\documentclass[12pt]{emulateapj}
\begin{document}
\title{The Growth \& Migration of Jovian Planets in Evolving Protostellar Disks with Dead Zones}
\author{Soko Matsumura\altaffilmark{1}}\affil{Department of Physics \& Astronomy,
Northwestern University, Evanston, IL, 60208 USA}

\author{Ralph E.~Pudritz}\affil{Department of Physics \& Astronomy, McMaster University, Hamilton, ON, L8S 4L8 Canada}

\and

\author{Edward W.~Thommes}\affil{Department of Physics, University of Guelph, Guelph, ON, N1G 2W1 Canada}

\altaffiltext{1}{This work was done while at McMaster University.}
%\author[Soko Matsumura et al.]{Soko Matsumura$^{1,4}$\thanks{E-mail:
%soko@physics.mcmaster.ca (SM); pudritz@physics.mcmaster.ca (REP); thommes@cita.utoronto.ca (EWT)},
%Ralph E. Pudritz$^{1,2}$ \&
%Edward W. Thommes$^{3}$\\
%$^{1}$Department of Physics and Astronomy, McMaster University, 1280
%Main Street West, Hamilton, ON, L8S 4M1, Canada \\
%$^{2}$Origins Institute, ABB 241, McMaster University, 1280
%Main Street West, Hamilton, ON, L8S 4M1, Canada \\
%$^{3}$Canadian Institute for Theoretical Astrophysics, University of
%Toronto, 60 St. George Street, Toronto, ON M5S 3H8, Canada \\
%$^{4}$Current location: Department of Physics and Astronomy,
%Northwestern University, 2145 Sheridan Road, Evanston, IL 60208-0834, USA}
%
%\def\LaTeX{L\kern-.36em\raise.3ex\hbox{a}\kern-.15em
%    T\kern-.1667em\lower.7ex\hbox{E}\kern-.125emX}
%
%\begin{document}
%
%\date{}
%
%\pagerange{\pageref{firstpage}--\pageref{lastpage}} \pubyear{}
%
%\label{firstpage}
%
%\maketitle
%
%--- abstract & keywords --------------------------------------------------------------------
%
\begin{abstract}
The growth of Jovian mass planets during migration in their
protoplanetary disks is one of the most important problems that
needs to be solved in light of observations of the small orbital
radii of exosolar planets.  Studies of the migration of planets in
standard gas disk models routinely show that the migration speeds
are too high to form Jovian planets, and that such migrating
planetary cores generally plunge into their central stars in less
than a million years.
%Recently, \cite{Alibert05} found that
%planetary migration has to be significantly slowed for the formation
%of Jovian planets. Here, we propose a possible mechanism which
%allows Jovian planet formation without slowing down migration
%artificially.
In previous work, we have shown that a poorly ionized, less viscous
region in a protoplanetary disk called a {\it dead zone} slows down
the migration of fixed-mass planets.
%we have pointed out the importance of dead zones in protostellar disks in significantly
%slowing the migration of planets of fixed mass.
In this paper, we extend our numerical calculations to include dead
zone evolution along with the disk, as well as planet formation via
accretion of rocky and gaseous materials.
%including an opacity reduction effect in an envelope of the protoplanet.
%as the planets migrate through the disk.
%a wide range of important evolutionary effects on planetary growth and
%migration including the evolution of disks with time as they accrete onto their central stars,
%the consequent shrinkage of disks dead zones with time, the accretion of
%rocky materials of the protoplanetary cores in a slower oligarchic phase, and the accretion
%of gas from the disks as the planets migrate through the disks.
Using our symplectic-integrator-gas dynamics code, we find that dead
zones, even in evolving disks wherein planets grow by accretion as
they migrate, still play a fundamental role in saving planetary
systems. We demonstrate
%by means of time-dependent simulations using our symplectic-integrator-gas dynamics code
that Jovian planets form within $2.5$ Myr for disks that are ten
times more massive than a minimum mass solar nebula with an opacity
reduction and without slowing down migration artificially.
%as in previous studies.
Our simulations indicate that protoplanetary disks with an initial
mass comparable to the minimum mass solar nebula (MMSN) only produce
Neptunian mass planets. We also find that planet migration does not
help core accretion as much in the oligarchic planetesimal accretion
scenario as it was expected in the runaway planetesimal accretion
scenario. Therefore we expect that an opacity reduction (or some
other mechanisms) is needed to solve the formation timescale problem
even for migrating protoplanets, as long as we consider the
oligarchic growth.
%rather than helping it in the oligarchic planetesimal accretion scenario.
%They are saved from splashing into their stars by the dead zones, which
%shrink with time from an intial radial extent of 13 AU to less than 2AU over a couple of million years.
%Thus, with all of these physical effect included, we find
%that dead zones, even in evolving disks wherein planets grow by accretion as they migrate,
%still play a fundament role in saving planetary systems.
We also point out a possible role of a dead zone in explaining
long-lived, strongly accreting gas disks.
\end{abstract}
%\begin{keywords}
%accretion, accretion disks - turbulence - planetary systems: formation
%- planetary systems: protoplanetary disks - planets and satellites:
%general - Solar system: formation - stars: pre-main-sequence
%\end{keywords}
%
%--- Section 1 ------------------------------------------------------------------------------
%
\section{Introduction}
The properties of nearly 300 recently discovered extrasolar
planetary systems reveal that Jovian mass planets are often found at
a scaled orbital radius of Mercury around their central stars
\citep[e.g.][]{Udry07}. Since none of the current theories of Jovian
planet formation can explain {\it in situ} formation of gas giants
at these small distances from their stars, it is generally agreed
that such planets were formed in the outer regions of protoplanetary
disks and migrated through them to their current positions
\citep{Lin96}. Growth of giant planets takes place during this
passage and both processes are terminated when most of the gas in
the protoplanetary disk is either accreted onto the central star, or
dissipated by photoevaporation
\citep[e.g.][]{Shu93,Hollenbach94,Alexander06b}. Observations of
infrared to submm emission and gas accretion rates onto the central
stars reveal that disk lifetimes are typically $1-10$ Myr
\citep[e.g.][]{Hartmann98,Muzerolle00,Andrews05,SiciliaAguilar06}.

The observed disk life-times raise two difficulties in planet
formation theory. The first regards planet migration: the migration
time scales of planets arising from the tidal interaction between a
planet and the gas disk are shorter than the disk lifetimes.
Therefore, unless they are stopped by some robust mechanism, planets
plunge into their central stars within about a million years.
%Why are there any planets at all?
The second is
%the accretion problem, which is
specific to core-accretion model for Jovian planet formation: the
formation time scales may be longer than, or comparable to, the disk
life times. Therefore, unless the formation time scale is reduced
somehow, the existence of giant planets cannot be explained by the
core accretion scenario.

The core-accretion model posits that gas giants result
from a two-stage process - the first being the formation of their
rocky cores by the repetitive coagulation and agglomeration of the
smaller bodies onto the larger ones (as in the terrestrial planet
formation), and the second being the accretion of their massive
gaseous envelope from the surrounding disk.
%The accretion problem is that this second stage takes an
%uncomfortably long time - comparable to the disk lifetime.
%
%Studies of the accretion problem go back to the pioneering work of
%\citet[][hereafter P96]{Pollack96} who performed numerical
%simulations of the growth of a protoplanet core with an initial mass
%of \(0.6 \ M_E\) by calculating the planetesimal and gas accretion
%rates in a self-consistent, interactive manner. They estimated that
%the {\it in situ} formation of Jupiter takes the uncomfortably long
%time of  \(8\times 10^6\).  A significant body of research has
%reduced this somewhat (see below), but this is still uncomfortable.

In the first stage, the formation of the rocky core proceeds as
bodies grow from a micron to a planetary size. One of the
difficulties occurs when particles grow up to dm sizes. At this
point, the collisional agglomeration may not be a preferred path to make the
larger bodies \citep{Langkowski08}.
%(Blum \& Wurm \citep{Blum05}).
Even if the sticking is still efficient and the bodies can keep on
growing, the gas drag becomes non-negligible for such objects. Since
the gas disk rotates at the sub-Keplerian speed, slightly slower
than these growing bodies, the bodies feel the head wind, lose the
angular momentum, and eventually migrate into the central star.
Migration induced by gas drag becomes most efficient for meter-sized
bodies, and becomes negligible again for the km-sized bodies since
they are large enough not to be affected by the gas drag
\citep{Weidenschilling77}.  The gravitational instability in the
planetesimal disk was suggested to avoid this meter-size barrier
\citep{Goldreich73}.  However, such a mechanism may be hindered by
the Kelvin-Helmholtz instability \citep{Cuzzi93,Weidenschilling95}
unless the local solid-to-gas ratio is sufficiently high
\citep[e.g.][]{Sekiya98,Youdin02}. Recently, it was demonstrated
that very rapid planetesimal growth up to Ceres-mass objects is possible
via a streaming instability in such turbulent disks
\citep[e.g.][]{Johansen07,Youdin07}. Therefore, there is good
justification to skip these early stages, and to assume that there
are already km-sized planetesimals as well as planetary embryo(s)
which are embedded in a gas disk.  In this paper, we will follow
this approach as the previous studies did.

%More problematic is the need to reduce the gas accretion timescale
%onto the rocky planetary core.
%Yet another difficulty occurs before a protoplanet achieves a
%crossover mass, above which gas accretion rate increases in a runaway fashion.
In the second stage, protoplanetary cores keep on accreting
planetesimals while they start developing gaseous envelopes. A
cornerstone work done by \cite{Pollack96} (P96 hereafter) identified
three phases in giant planet formation. The first phase is a rapid
core building phase, in which a protoplanetary core of $0.6 M_E$
grows up to \(\sim 10 M_E\) within half a million years.  The second
phase is a slow gas and planetesimal accretion phase which lasts
until the crossover mass (for which the envelope mass is comparable
to the core mass) is reached. The third phase is a rapid gas and
planetesimal accretion phase, in which the planet quickly becomes a
gas giant.  P96 showed that planet formation spends most time in the
second phase that could last for several million years.  Since
this is uncomfortably close to a typical disk life time, it was
considered as one of the weakest points of core accretion scenario
\citep[e.g.][]{Boss97}. To shorten the second phase, protoplanets could
either increase the gas accretion rate itself, or increase the
planetesimal accretion rate and expand the gas feeding zone.
%
%One approach is to more carefully consider the mechanisms that could
\cite{Ikoma00,Hubickyj05} (hereafter INE00, and HBL05 respectively)
took the former approach, and considered a reduced opacity in planetary envelopes.
%\citep[][hereafter INE00, and HBL05 respectively]{Ikoma00,Hubickyj05}.
HBL05 used an updated version of the code by P96 with a more recent
opacity table, and showed that smaller opacity, which mimics the
dust settling and coagulation in the protoplanet's atmosphere,
decreases the effective gas pressure of the protoplanetary envelope,
and hence leads to the faster gas accretion. They showed that the
{\it in situ} formation timescale of Jupiter could be as short as 1 Myr
if the opacity is \(2\%\) of the interstellar value
and the core mass is \(10 M_E\).
%
%Another factor that could in principle reduce the gas timescale is
%that gas accretion occurs during migration \citep[][hereafter
%A05]{Alibert05}.
On the other hand, \cite{Alibert05} (hereafter A05) took the latter
approach, and considered planet formation during migration.  In P96,
planetesimal accretion slows when the planetesimals get depleted in the
feeding zone, because
%before the crossover mass for rapid gas accretion is
%reached. Then
a planet has to accrete gas and expand its feeding
zone to further accrete planetesimals.
Since the gas accretion in Phase 2 tends to be slower than the planetesimal accretion
in Phase 1, it takes longer to accrete a similar amount of gas envelope to the core,
and achieve a crossover mass.
A05 overcame this problem by including disk evolution and planet migration
in the model, so that the planetesimal feeding zone is constantly replenished.
The gas accretion time shortens as the planetary mass increases, because
it proceeds on the Kelvin-Helmholtz timescale, which is a steep
function of planetary mass.
They showed that the Jupiter formation timescale could be of the order of 1 Myr
if the planet migrates.
%Since the gas accretion proceeds on the Kelvin-Helmholtz timescale, which is a steep
%function in planetary mass, the gas accretion time shortens as the planetary mass increases.
However, as noted by \cite{Ward97}, solving
the accretion problem by migration in a standard disk model is a
``double-edged sword", since it comes at the price of possibly
losing the planets to their central star. Thus, A05
%did not address this problem and simply
assumed that some unspecified mechanism would slow down the fast
type I (pre-gap opening) migration by factors of \(10-100\) times
compared to the migration estimated in a 3D disk by \cite{Tanaka02}.
%slower than the migration rate estimated for a
%3D disk by \cite{Tanaka02}.

%The disk models that were employed in the discussion above all
%assume that a protostellar disk is turbulent throughout its entire
%radial extent.
Here, we propose a possible mechanism which can slow planet
migration, and investigate planet formation in that context.
In a standard disk with a smooth surface mass density and
a standard viscosity expected from the MRI turbulence, both type I
and II migration time scales are shorter than the disks' life times.
%Disk accretion onto the central star is driven by viscosity in a
%differentially rotating disk.
The most popular source of such a
viscosity is the magneto-rotational instability \citep[MRI,][]{Balbus91}.
However, it has been demonstrated by several groups that
protoplanetary disks are not MRI active everywhere, but harbor
extended regions known as the dead zones \citep{Gammie96} where
there is virtually no turbulence
%%%%%%%%%%%%% Foot Note
\footnote{This arises in principal because operation of the  MRI
requires good coupling between the gaseous disk and the magnetic
field.  However, the ionization fraction in the inner regions of dense
protoplanetary disks is typically so low that Ohmic
diffusion prevents the growth of the MRI in such regions. The size
of the dead zone can be computed once the source of disk ionization
(e.g. stellar X-rays, cosmic rays) is known.  Assuming that the MRI
is the dominant source of the disk's ``turbulent'' viscosity, the
dead zone is nearly inviscid. The existence of a dead zone in a
protoplanetary disk was first proposed by \cite{Gammie96} who showed
that there is a magnetically dead zone in the inner disk where
cosmic rays cannot penetrate (assuming the cosmic ray stopping
density is \(98 \ {\rm g \ cm^{-2}}\)), and that such a region is
sandwiched by the magnetically active surface layers where gas
accretes toward the star efficiently.}.
%%%%%%%%%%%%%% Foot Note END
Many authors have studied the physical extent of the
dead zones
%has been studied by many authors
by using different disk models and taking account of a variety of
the ionization/recombination processes \citep[e.g.][hereafter MP03,
and MP06
respectively]{Glassgold97,Sano00,Fromang02,Semenov04,Matsumura03,Matsumura06}.
These models suggest roughly the same size of the dead zones (from
\(<1\) AU to \(10-20\) AU), indicating that a dead zone is a robust
feature of a protoplanetary disk. Since this is a critical region of a
disk both for planet formation and migration, it is of vital
importance to investigate these problems in the context of
protoplanetary disks with dead zones.
Recent numerical simulations of such a layered disk have shown that
the expected viscosity parameter \(\alpha\) is about \(10^{-2}\) in
the active zone and \(10^{-4}-10^{-5}\) in the dead zone
\citep{Fleming03}. Therefore, if the MRI is the dominant source of
the disk viscosity, a significant decrease in viscosity is expected
within the dead zone, which most likely affects the rate of planet
migration. Moreover, recent observations have revealed that disk's
viscosity parameters may take a larger range than previously
expected; \(\alpha=10^{-6}-10^{-1}\) \citep{Hueso05,Luhman07},
indicating that at least some observed disks may possess a low
viscosity region like a dead zone.

%Of central importance for planet formation and migration is the fact
%that dead zones can significantly slow the rate of migration of
%planets.
In an earlier paper, we have demonstrated that dead zones can
significantly slow the rate of migraion of planets, and save
planetary systems from plunging into the central stars
\citep[][hereafter MPT07]{Matsumura07}.   In particular, we found
(1) type II migration (post-gap-opening migration) is slowed in the
dead zone due to the low viscosity there, (2) even low-mass planets
(\(\leq 10 M_E\)), which are usually type I migrators (i.e. non
gap-openers), may open a gap in the dead zone if the thermal
condition is satisfied, and thus migrate slower there, and (3) type
I migrators moving toward the dead zone can be stopped at the outer
edge of the dead zone due to the jump in mass density there
%at the outer edge of the dead zone,
%increase inside it,
which is a result of the slower advection speed inside the dead zone with respect to outside it.
%gas accretion inside the dead zone
%with respect to outside it.

%This is because one of the difficulties of planet formation is that the
%migration timescale is much shorter compared to the formation
%timescale, and thus growing planets are prone to being lost into the
%central stars.

In this paper, we present a rather complete treatment of the
formation and migration of Jovian planets within their evolving
protoplanetary disks.
%The mass evolution of the disk is a
%consequence of viscous accretion which enables the transport of most
%of the disk material into the central star in roughly 10 million
%years.
%(we end with a Jovian mass disk typically).
We compute comprehensive time-dependent models for the growth of
Jovian planets in the core accretion picture in viscously evolving
gaseous disks.  Our calculations include updates on both stages of
accretion (planestimals and gas), as well as the ability to follow
planetary migration down to 0.1 AU.
%for up to 10 million years.
We approach this problem by means of time-dependent simulations that
track both planetary accretion and migration through disks with
evolving dead zones.
%An important aspect of dead zone evolution that
%should be taken into account is its
There are two major effects of disk evolution in our models: (i) the
gradual accretion of mass from the disk onto the star which limits the
reservoir that is available to the growing Jovian planet; and (ii)
the shrinkage of the size of the dead zone as a consequence of mass
accretion onto the star (mainly) through the well
coupled surface layers of the disk. This process reduces the size of
the dead zone substantially with time --- perhaps leaving it only a
few AU in extent after a million years.
%
%Our major findings are (1) when a planetary core starts its
%migration and growth from outside the dead zone, the core is kept
%outside the dead zone due to a surface mass density jump and grows
%up to a Jupiter mass, and (2) when a planetary core starts from
%inside the dead zone, the core is left outside the dead zone which
%shrinks due to mass accretion through surface layers, and grows up
%to a Jupiter mass. Jovian planets in both of these circumstances are
%accreted within \(7\times 10^6\) years.  We require models that are
%initially an order of magnitude greater in column density than the
%minimum mass solar nebula.
%
%, and (3) when a
%planetary core starts well inside the dead zone, the core grows
%inside the dead zone, opens a gap, and stops its gas accretion.

%the planet migration and formation by taking account of the dead zone.
%--- poorly ionized region in a protostellar disk which
%has no magneto-rotational instability (MRI) turbulence
%\citep{Gammie96}, and hence expected to be nearly inviscid.
We first introduce our disk models and numerical methods in \S 2. We
also highlight a possible role of dead zones in disk accretion onto
the central stars. Then we study planet formation in an inviscid
disk, and compare our results with P96, and HBL05 (\S 3). We also
check the effect of the opacity, and compare the results with HBL05,
which improved the work of P96 and further investigated the opacity
effects. We then go on to compute planet formation and migration in
an evolving disk, and compare the results with A05 in \S 4. The
culmination of our work is presented in \S5 where we generalize our
results and present planet formation in an evolving disk with a dead
zone. Finally, we summarize our work in \S 6.
%
%--- Section 2 ------------------------------------------------------------------------------
%
\section{Numerical methods and Disk models}
In this section, we introduce the numerical code that we use to
simulate planet formation in a realistic protoplanetary disk. We
focus on the newly added features and refer the reader to our
previous paper (MPT07) for details of the basic code.  We also
summarize the initial conditions chosen for the runs in the
following sections.

We perform numerical simulations of planet formation and migration
by using a hybrid numerical code, which combines an N-body
integrator with a simple disk evolution code \citep{Thommes05}. The
N-body part is based on the symplectic integrator called SyMBA
\citep{Duncan98}, which has improved the N-body map of
\cite{Wisdom91} with an adaptive timestep to handle the close
encounters among massive bodies.  The gaseous disk part of the code
evolves a disk viscously as well as through angular momentum
exchange with the embedded planets according to a general
Navier-Stokes equation. By following the standard prescription by
\cite{Lin86}, the gas disk is divided into radial bins, which
represent annuli of disks with azimuthally and vertically averaged
properties like surface mass density, temperature, and viscosity.
Viscous evolution of the disk is calculated by specifying the
standard alpha viscosity \citep{Shakura73} for each bin, while the
effect of disk-planet interactions is added in the form of the
torque density as in \cite{Ward97,Menou04}.

The code is also modified to simulate planet formation, which
consists of planetesimal and gas accretion. 
For planetesimal accretion, 
as it was mentioned in \S 2, the transition between runaway, and 
oligarchic phases occurs once the
protoplanet becomes locally large enough to perturb the surrounding
smaller planetesimals and increase their random velocities, which
decreases the collisional crosssection, and therefore leads to the
longer accretion time \citep{Ida93,Kokubo98}.  This critical mass is
about a few times $10^{-3}$ Moon masses or less \citep{Thommes03}.
Since all of our simulations start with \(0.6 \ M_E\), we can safely
assume that the oligarchic growth is the dominant planetesimal
accretion phase in our case.
%and that the growth
%is much slower compared to the runaway growth which has been adopted
%in major works of planet formation (e.g. P96, HBL04, and A05).
%\cite{Thommes03} studied the oligarchic growth of \(10\) km-sized
%planetesimals in a disk which is 10 times more massive than the
%minimum mass solar nebula model, and found that making a 10 Earth
%mass core at 5 AU takes as long as the disk's lifetime (\(\sim
%10^7\) years).

The planetesimal 
accretion part of the code was developed by \cite{Thommes03} to 
handle the {\it oligarchic} growth, which is the slower planetesimal
accretion phase following the rapid planetesimal accretion
\citep{Ida93,Kokubo98}. Here, we follow the previous studies
\citep[e.g.][]{Kokubo96,Thommes03}, and assume (1) dynamical
friction by smaller planetesimals on larger ones is effective, so
that larger planetesimals have smaller velocities than smaller ones,
and (2) gravitational focusing is effective, so that the relative
velocities between two colliding bodies are smaller than the escape
velocities from the larger ones, and therefore collisions
effectively lead to coagulation. Accretion rate of planetesimals
with mass $m$ onto a protoplanet with mass $M_p$ is written as
\begin{equation}
\frac{dM_p}{dt}\sim \frac{ C\Sigma_{solid} M_p^{4/3} }{ e_m^2 a^{1/2} } \
,
\end{equation}
where $C=6\pi^{2/3}[3/(4\rho_M)]^{1/3}(G/M_*)^{1/2}$, $\rho_M$ is
the bulk density of a protoplanet, $\Sigma_{solid}$ is the surface
mass density of a planetesimal disk, and $e_m$ is the eccentricity of
planetesimals. We assume that the protoplanet accretes planetesimals
within its feeding zone of $10R_{\rm Hill}$ ($R_{\rm
Hill}=(M_p/(3M_*))^{1/3} a$ is the Hill radius of the protoplanet),
which is a typical orbital separation between protoplanets \citep{Kokubo98}.

The gas accretion part of the code estimates the planetary mass
increase by calculating the gas accretion timescale and radius (see
\S 2.1 for details).  Instead of solving the planetary structure
equations directly to estimate the gas accretion rate as in P96,
HBL05, and A05, we calculate the gas accretion timescale based on
these previous studies, and let the protoplanetary core accrete gas
inside the accretion radius on the estimated timescale.
%The protoplanetary core simply accretes gas
%inside the accretion radius on the estimated accretion timescale.
%Therefore, different from P96, HBL05, or A05, we are not calculating
%the planetary structure to estimate the gas accretion rate.  {\bf
%want to include why this is justified?}

We also add a new capability to our disk model to include the
evolution of a dead zone due to the faster accretion through the
well coupled surface layers onto the central star. This is further
explained in \S 2.2.
\subsection{Gas accretion prescription}
Since the planetesimal accretion prescription is discussed in detail
elsewhere \citep{Thommes03}, we focus on the gas accretion
prescription added to the code in this subsection.
%Following the previous studies, we define three stages of the gas
%accretion; (1) gas accretion while planetesimal accretion is
%on-going, (2) gas accretion after planetesimal accretion ceases,
%and (3) gas accretion through the subdisk around the planet.
%
Gas accretion onto a protoplanet has been studied either by solving
the planetary structure equations (e.g. P96, HBL05, INE00, and A05),
or by performing the hydrodynamic (HD) simulations
\citep[e.g.][]{Bryden99,Lubow99,Kley01,DAngelo02,Tanigawa02,DAngelo03,Bate03}.
The main focus of the former studies is on the core-building phase
to the rapid gas accretion phase, while that of the latter ones is
on the final stage of gas accretion where a massive planet accretes from its
subdisk.
Since the former studies except A05 treat the gas disks rather
simply by assuming that the protoplanetary feeding zone is always
replenished, they have a relatively crude estimate of the available
amount of gas in the feeding zone. Therefore, they tend to
overestimate the gas accretion for a planet which is massive enough
to open a gap.  The gas accretion in the latter case is regulated by
the subdisk formed around the planet. Since these studies do not
include the contraction of the protoplanetary atmosphere, they tend
to overestimate how quickly a protoplanetary core can actually
accrete gas. Here, we try to combine the strengths of these two
methods, and use three different gas accretion rates depending on
presence/absence of planetesimal accretion, and circumplanetary disk
accretion.
%
%*** SOKO: [ I don't think all of those lines need to be put into the Figure.  At a minimum,
%need the 3 basic curves (equ. 1, 2, 3).  Leave description of other papers in text, but
%just show your model.  (perhaps could put in the fig. some idea of the "scatter" that
%is implied by the totality of the models]
%
%Fig. \ref{fig1} presents a compilation of gas accretion timescales
%using data obtained from different studies. Dotted line is gas
%accretion timescale estimated from P96, while a similar study done by
%\cite{Tajima97} is shown in the light grey descending line.
%Two parallel descending
%lines are gas accretion while planetesimal accretion is on-going
%(upper line) and gas accretion after planetesimal accretion ceases
%(lower line) respectively (INE00, see Eq. (1) and (2).)
%These descending trends
%predict shorter accretion timescales for larger planets since the
%models don't take account of disk evolution effects. The
%parabola-like relation is based on the HD simulation by
%\cite{DAngelo03}, which predicts much longer gas accretion timescale
%for larger planets due to gap opening.  Similar study done by
%\cite{Tanigawa02} is shown in the lowermost light grey line. Since
%this model does not include a gap-opening effect, it shows the
%``true'' accretion timescale of a planet.  Possible gas accretion
%timescales throughout planet formation are shown in solid lines.

\begin{figure}[b]
%\begin{sideways}
%\includegraphics[angle=-90,width=8.5cm]{figs3/comp_accr_time.eps}
%\includegraphics[width=8.5cm]{figs4/comp_accr_time2.eps}
%\plotone{figs4/comp_accr_time3.eps}
\plotone{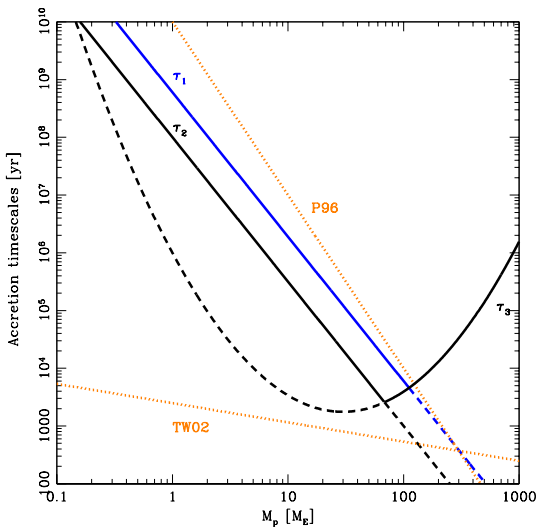}
%\end{sideways}
\caption[fig1]{Comparison of gas accretion timescales as a function
of planetary mass. The descending solid and the upper dotted line
are the gas accretion timescales for the envelope contraction phase,
while a parabola and the lower dotted line are those for the subdisk
accretion phase. The upper dotted line is the gas accretion rate
estimated from the simulation of P96. Blue line (the upper of the
parallel lines) is gas accretion while planetesimal accretion is
on-going, while black one (the lower of the parallel line) is that
after planetesimal accretion ceases. Both of them are based on the
study by INE00.  Black parabola and the lower dotted line are the
subdisk accretion timescales from \cite{DAngelo03} and TW02,
respectively.
%The lower dotted line is
%%the envelope contraction phase, and
%the accretion rate during subdisk accretion phase taken from \cite{Tanigawa02}.i
The subdisk accretion by \cite{DAngelo03} appears to be slower than
TW02 since the gap-opening effect of a planet is included.
\label{fig1}}
%
%The solid lines show the cases with the standard opacity (\(\kappa=1 \ {\rm cm^2 \
%g^{-1}}\)) while the dotted ones show the cases with a reduced
%opacity (\(\kappa=0.03 \ {\rm cm^2 \ g^{-1}}\).  The ascending lines
%including the parabola are the gas accretion times for the subdisk
%accretion phase. \label{fig1}}
\end{figure}

The presence, or absence of planetesimal accretion affects the
efficiency of gas accretion.
%As stressed in P96, the efficiency of
%gas accretion depends on the rate of planetesimal accretion.
INE00 suggested that, when the planetesimals are accreting
concurrently with the gas, the thermal energy released in the
bombardment of planetesimals provides extra pressure support for the
gaseous envelope of the protoplanet, and therefore the gas accretion
slows down. They found that such an accretion rate could differ from
the accretion rate without planetesimal accretion by several times,
which are plotted in Fig. \ref{fig1} as two parallel lines.
%The effect can be seen in the dark and grey lines in Fig. \ref{fig1}.
They also studied the effects of core mass and dust opacity on gas
accretion, and showed that there is an optimal core mass which
results in efficient gas accretion. If planetesimal accretion is cut
off long after the protoplanet starts accreting gas actively, the
gas accretion time becomes longer. This is because the radiative
loss is mostly compensated by planetesimal accretion rather than gas
accretion (i.e. gravitational energy release due to the envelope
contraction). On the other hand, if planetesimal accretion is cut
off while the core is still small, the gas accretion time is also
prolonged, because the smaller mass leads to the smaller radiative
loss and hence the slower envelope contraction.

The final stage of gas accretion is likely to be controlled by the
subdisk around a planet \citep{Tanigawa02,Bate03,DAngelo03}.
%studied the gas accretion
%rate onto a protoplanetary core by performing two-dimensional HD
%simulations.
%They have argued that there is a critical mass above
%which the gas accretion is controlled by the subdisk around a
%protoplanet.
Gas accretion through subdisks becomes less efficient for more
massive planets because the planets open a wider and deeper gap
\citep[e.g.][]{Bate03}. The effect is seen in the parabola-like
curve in Fig. \ref{fig1} by \cite{DAngelo03}. This, however, is not
clear in \cite[][hereafter TW02]{Tanigawa02}'s case (the lower
dotted line), because they don't take account of a gap-opening
effect. In other words, their gas accretion rate is the ``raw''
accretion rate of the planet. Below the critical mass for subdisk
accretion, gas accretion is expected to be controlled by the
contraction of protoplanetary atmosphere. Therefore, we apply the KH
timescale for small protoplanets, and subdisk accretion timescale
for larger protoplanets.

Following these studies, we define three stages of the gas
accretion; (1) gas accretion while planetesimal accretion is
on-going, (2) gas accretion after planetesimal accretion ceases, and
(3) gas accretion through the subdisk around the planet. Fig.
\ref{fig1} presents a compilation of these gas accretion timescales
obtained from different studies.
%Dotted line is gas
%accretion timescale estimated from P96, while a similar study done by
%\cite{Tajima97} is shown in the light grey descending line.
Two parallel descending lines are gas accretion time scales while
planetesimal accretion is on-going (upper line) and after it ceases
(lower line) respectively (INE00, see Eq. (1) and (2)). For
comparison, accretion time scales obtained from P96
%and a similar study by \cite{Tajima97} are
is shown in the upper dotted line. These descending trends predict
shorter accretion timescales for larger planets because the models
don't take account of disk evolution effects nor planetary subdisks.
On the other hand, the parabola-like relation is based on the HD
simulation by \cite{DAngelo03}, which predicts much longer gas
accretion timescale for larger planets due to gap opening and
accretion via circumplanetary disks. A similar study done by TW02
without a gap-opening effect is shown in the lower dotted line.
%Since this model does not include a gap-opening effect, it shows the
%``true'' accretion timescale of a planet.
Possible gas accretion timescales throughout planet formation are
shown in solid curves.
%
%In this study, we will take account of the above effects and adopt
%the three different accretion timescales, which are shown in Figure 1.
%
%
%**** SOKO: Different rates OK – but need to show the net result of all
%3 regimes in one resulting curve that covers all regimes and is
%plotted in Figure 1.  Not clear that we need to plot all the other
%curves from other people. Fig 1 still too confusing to use by
%general reader. *****

(1) When there is planetesimal accretion as well as gas
accretion, we follow the timescale by INE00 (the upper line of the
solid parallel lines):
\begin{equation}
\tau_{1}=6.\times10^8\left(\frac{M_p}{M_*}\right)^{-2.5}
\left(\frac{\kappa}{\rm 1 \ cm^2 \ g^{-1}}\right) \label{tau1} \ ,
\end{equation}
where \(\kappa\) is the dust opacity in the planetary envelope.

(2) When the planetesimal accretion ceases, but the gas accretion is
still on-going, we adopt the timescale by INE00 (the lower line of
the solid parallel lines):
\begin{equation}
\tau_{2}=1.\times10^8\left(\frac{M_p}{M_*}\right)^{-2.5}
\left(\frac{\kappa}{\rm 1 \ cm^2 \ g^{-1}}\right) \label{tau2} \ .
\end{equation}
Both of the above equations show shorter accretion timescales as the
planetary mass grows.

(3) When a protoplanet becomes large enough to have a subdisk, the
gas accretion timescale can be described by
\cite{DAngelo03}:
\begin{eqnarray}
\tau_{3}&=&\frac{M_p}{\dot{M}_p} \\
\log\left(\frac{\dot{M}_p}{M_E \ {\rm yr^{-1}}}\right) &=&
\left(18.47+9.25\log\left(\frac{M_p}{M_*}\right)+1.266\log\left(\frac{M_p}{M_*}\right)^2\right)
\label{tau3} \ .
\end{eqnarray}
The transition of the first to the second phase occurs when the core
accretion rate drops to zero. The transition of the second to the
third phase occurs when \(\tau_2\sim\tau_3\). For a standard opacity
\(\kappa\sim 1 \ {\rm cm^2 \ g^{-1}}\), the crossover mass is about
\(70 \ M_E\) as it can be seen in Fig. \ref{fig1}.

A protoplanet accretes gas within its accretion radius on each of
these time scales. As in P96, the accretion radius for Eq.
\ref{tau1} and \ref{tau2} is set to either Hill ($R_{\rm
Hill}=(M_p/(3M_*))^{1/3} a$), or Bondi ($R_{\rm Bondi}=GM_*/c_s^2$,
where $c_s$ is sound speed) radius, whichever is smaller. On the
other hand, the accretion radius for Eq. \ref{tau3} is chosen to be
$2 R_{\rm Hill}$, which is motivated by numerical studies like
\cite{DAngelo03}.

\subsection{Accretion histories in disks with dead zones}
Although our disk model is one-dimensional, it is still possible to
include the effects of azimuthal and vertical structure through disk
parameters in an averaged way. The difference in disk evolution due
to a vertical structure becomes particularly important for a disk
with a dead zone. Generally, a dead zone is sandwiched between the
upper and lower turbulent surface layers where the disk is
well-ionized \citep{Gammie96} and hence the MRI is active.
Therefore, the mass accretion is expected to be more efficient in
active layers than in the dead zone.  In MPT07, we did not include
this effect, and hence found very little decrease in disk mass even
after \(10^7\) years.

Here, we take account of the effects of disk's vertical structure in
the following simple way. The mass accretion through the active
layers and the dead zone toward the central star can be written as
the summation of accretion rates in the active layers and the dead
zone \citep[e.g.][]{LyndenBell74,Gammie96,Fromang02}:
\begin{equation}
\frac{dM}{dt}= 6\pi r^{1/2} \frac{\partial}{\partial r}(2\Sigma_{\rm crit}
\ \nu_{\rm active} r^{1/2} + (\Sigma-2\Sigma_{\rm crit}) \
\nu_{\rm dead} r^{1/2}) \ ,
\end{equation}
where \(\Sigma_{crit}\) is the critical surface mass density below
which the disk is well-ionized, while \(\Sigma\) is the total gas
surface mass density. Specifically, we assume that the local disk is
``dead'' if \(\Sigma > \Sigma_{crit}=21\), and \(80 \ {\rm g \
cm^{-2}}\) for the surface mass density at 1 AU of 
\(\Sigma_0=10^3\) and \(10^4 \ {\rm g \ cm^{-2}}\),
respectively. The factor of 2 in front of \(\Sigma_{\rm crit}\)
implies the upper and lower layers. Also, \(\nu=\alpha c_s h\) is
the viscosity (\(\alpha\) is the viscosity parameter, \(c_s\) is the
sound speed, and \(h\) is the disk pressure scale height), and the
subscripts dead and active imply parameters in the dead zone and
active layers respectively.

From the above equation, we can define the ``vertically averaged''
disk viscosity parameter that is weighted by the surface mass
density:
\begin{equation}
\alpha=\frac{1}{\Sigma}\left(2\Sigma_{\rm crit}\alpha_{\rm active}+
\left(\Sigma-2\Sigma_{\rm crit}\right)\alpha_{\rm dead}\right) \label{avealpha} \ .
\end{equation}
In our simulations, the protoplanetary disks evolve viscously according to this viscous $\alpha$.

\begin{figure}
%%\includegraphics[width=12cm]{figs_chap6/test911more_surf.eps}
%\includegraphics[width=7cm]{figs_chap6/test911more_surf.eps}
%\includegraphics[width=7cm]{figs3/loglog1e7.eps}
%\plotone{figs3/loglog1e7.eps}
\plotone{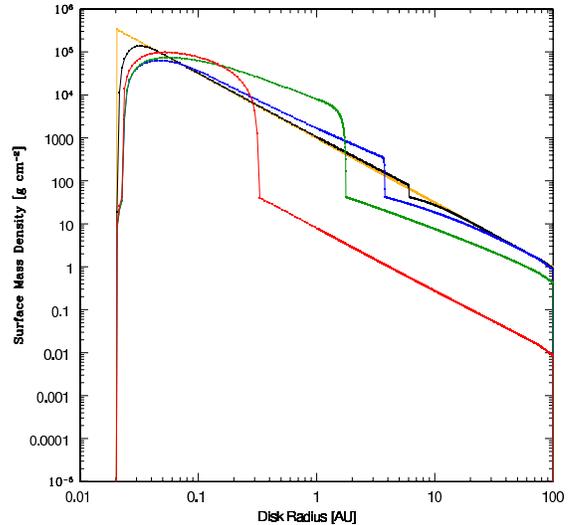}
\caption[Surface mass density evolution of a disk with a dead
zone]{The surface mass density at \(t=0, \ 10^4, \ 10^5, \ 10^6\),
and \(10^7\) years for disk with dead zone (DZ) but no planet (RunA).
The dead zone shrinks as the disk evolves.
%The rough features seen in the figure arise partially due
%to the fact that we cut off the surface layer accretion at the
%outer dead zone radius.
\label{fig2}}
\end{figure}

\begin{figure}
%\includegraphics[width=7cm]{figs3/disk_1e7.eps}
%\epsscale{2}\plottwo{figs4/disk_2e7.eps}{figs4/diskaccr_diskmass_2e7.eps}
\epsscale{2}\plottwo{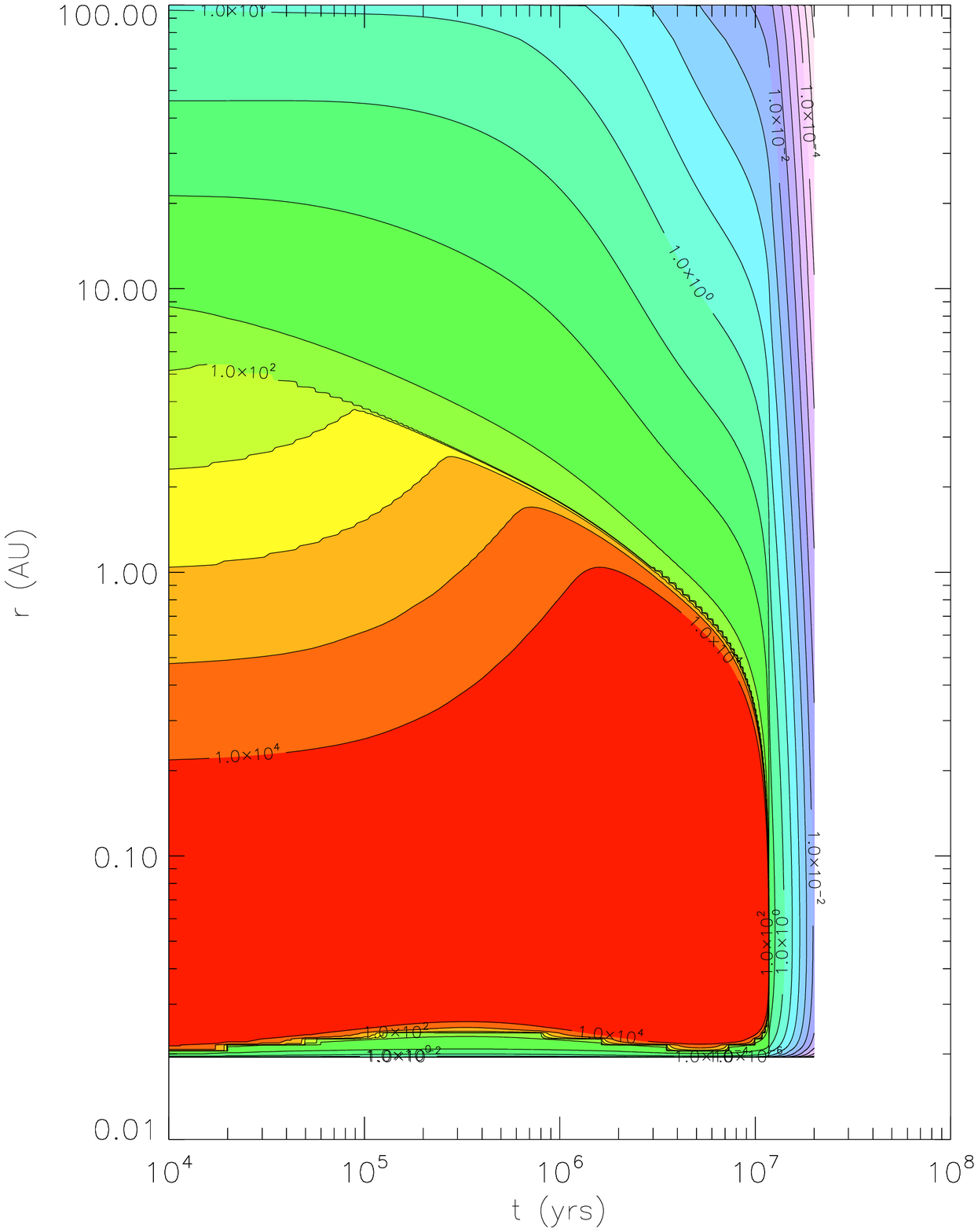}{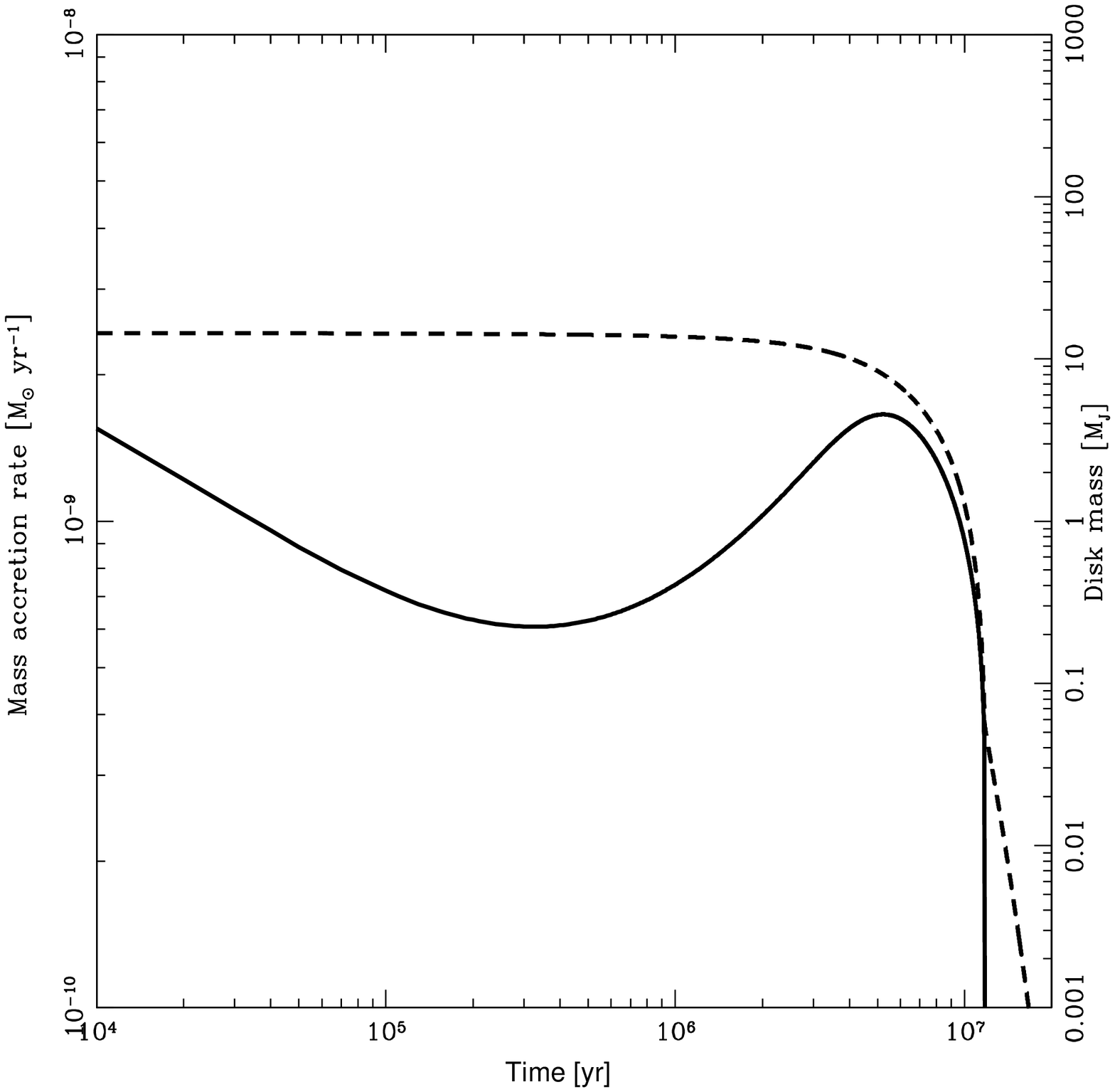}
%\unitlength1cm
%\begin{minipage}[t]{9cm}
%%\hspace{0.5cm}
%\begin{minipage}[h]{9cm}
%\begin{picture}(1,6)
%\includegraphics[width=4.5cm]{figs4/disk_2e7.eps}
%\end{picture}
%\hspace{3.5cm}
%\begin{picture}(7,6)
%\includegraphics[width=4.5cm]{figs4/diskaccr_diskmass_2e7.eps}
%\end{picture}
%\end{minipage}
%\end{minipage}
%\begin{minipage}[t]{9cm}
%%\vspace{0.5cm}
%
\caption[Evolution of a disk with a dead zone]{Evolution of
disk with DZ but no planet (RunA) Top: Evolution of
the disk column density. The disk
viscosity parameter is set to \(\alpha=10^{-5}\) inside the dead
zone, and \(10^{-2}\) outside it. The dead zone shrinks rapidly due
to mass accretion through the surface layers. Bottom: Corresponding
plots of disk accretion rate (solid line) and disk mass (dashed
line). Once the dead zone is gone, the disk disappears quickly.
\label{fig3}}
%\end{minipage}
\end{figure}

Now we will show one example which describes a typical evolution of a disk with a dead zone.
Using the averaged alpha value in Eq. \ref{avealpha}, 
and assuming that \(\alpha_{\rm
active}=10^{-2}\) and \(\alpha_{\rm dead}=10^{-5}\), we evolve 
a MMSN-type disk with \(\Sigma=10^3 (a/AU)^{-3/2} \ {\rm g \, cm^{-2}}\) 
as in Fig. \ref{fig2} and \ref{fig3} (RunA,
see Table \ref{tb1} for initial conditions).
Fig. \ref{fig2} shows the surface mass density evolution of a
disk with a dead zone, including the effect of the surface layer
accretion. There is no planet in this case.
An important feature of disks with dead zones is immediately
apparent in this figure.  The high advection speed in the outer
disk, compared to the much lower speed in the region containing the
dead zone, results in a pile up of gas. This
manifests itself in a steep density gradient, which we have already
shown can play an important role in reflecting the inwardly migrating
planets in the outer disk regions (MPT07).  As material accretes
along the surface of the dead zone, we see that this dense front
moves inward.
The dead zone shrinks
very rapidly, and only within 2 AU is dead by 2 Myr.
We will see how the evolution of a dead zone affects planet formation
in \S 5.
%We assume here of course, that the disk has a finite reservoir of
%mass which is gradually deposited onto the central star.
%The surface mass density becomes nearly constant inside the dead zone.
%
%The disk surface mass density forms a much
%smaller surface mass density jump compared to MPT06, and the dead
%zone shrinks rapidly --- to within 2 AU is dead by \(2 \times 10^6\)
%years.

The corresponding figure for the disk evolution is shown in the top
panel of Fig. \ref{fig3}.
Note that the dead zone edge is where the surface
mass density contours are the densest.
The accretion onto the central star
speeds up as the dead zone disappears. The evolution of mass
accretion rate (solid line) and disk mass (dashed line) for the same
run are shown in the bottom panel of Fig. \ref{fig3}.
With a dead zone, the mass accretion rate is nearly constant 
for a few Myr, and more than $10\%$ of the initial disk mass accretes 
onto the central star once the dead zone is gone at around 10 Myr.
This final dispersal of a disk proceeds at the viscous time scale of 
a turbulent disk if there is no other mechanism, and takes about a few Myr.
%
%The total disk mass is about \(0.01\) solar mass initially,
%but is less than a Jupiter mass by \(10^7\) years, which roughly
%corresponds to the mass in transition disks {\bf check this
%carefully + Ref: Kamp, Fraudlling, \& Chengalur 2007}.
%Fig. \ref{MPT06} is the corresponding figure for the evolution of
%such a disk.

This characteristic evolution of the disk accretion rate in disks
with dead zones may have an interesting implication for the
observations. The current observations indicate that mass accretion
rate is a gradually decreasing function in time
\citep[e.g.][]{SiciliaAguilar06}, although the range in mass
accretion rate $10^{-10}-10^{-7} \ {\rm M_{\odot} \, yr^{-1}}$ is
roughly constant between $1-10$ Myr.  
%
%Also, the final dispersal of
%a gas disk is a rather fast ($\sim 10^5 \ {\rm yr}$) process. 
%Such a ``two-time-scale'' behavior cannot be explained only by viscous 
%evolution of a standard disk, nor a disk with a dead zone, but is 
%well-explained by considering viscous evolution of a disk along 
%with photoevaporation
%\citep{Clarke01,Alexander06b}.  Once the viscous accretion rate
%becomes comparable to the wind mass loss rate, the evaporated amount
%of mass cannot be compensated by viscous disk accretion, and
%therefore leads to the division of the inner and outer disks. The
%inner disk accretes quickly (a few $10^4$ yr) to the central star,
%while the outer disk evaporates depending on the strength of the
%stellar flux \citep{Alexander06b}.  
%If this is the case, disk accretion rate is expected to decrease 
%gradually over time (see Fig.
%\ref{fig6} for example) until the photoevaporation kicks in. In such
%a scenario, 
In a standard disk with no dead zone, a gradual decrease of disk mass 
is expected (see Fig. \ref{fig6} for example), and therefore 
we may need very massive initial disks to explain the long-lived, 
strong accretors.

Alternatively, if the dead zone is playing a role in disk dispersal,
the mass accretion rate is likely to stay nearly constant throughout
the evolution until the dead zone disappears.  In this case, mass
accretion rates correlate with disk masses as well as the difference
between effective viscosities inside and outside the dead zones.
Once the dead zone disappears, the rest of the disk accretes onto
the central star at a higher viscous accretion rate.
%, unless there is some 
%other mechanism like photoevaporation, which removes disk on a 
%shorter time scale of a few $10^4$ \citep{Alexander06b}.  
%Unless photoevaporation
%kicks in, this final dispersal of a disk takes about a few Myr.
Currently, it is difficult to say whether gas accretion rates tend to
decrease gradually, or stay rather constant for a significant part of the
disks' lifetime.
%It is difficult to distinguish these two evolutionary histories of mass
%accretion rates currently.
Future observations may constrain masses
and accretion rates of protoplanetary disks more precisely to
indicate a possible preferred path.
\subsection{Disk models}
We study planet formation in a protoplanetary disk by assuming that
a single planetary embryo with \(0.6 \ M_E\) is embedded in gas and
planetesimal disks.  This is the same initial core mass used in P96,
HBL05, and A05.
%The disk models are consist of the gas and planetesimal disks.
In this paper, we use three slightly different gas disk models to
compare our results directly with the previous studies, but keep the
solid surface mass density the same
\(\Sigma_{solid}=270(r/AU)^{-2}\) for all the models. Instead of
choosing a specific planetesimal size, we use different planetesimal
sizes ranging from 100 m to 100 km, and compare their results with
one another.
%
%We use three slightly different gas disk models.
%This is because we would like to compare our results directly with those by
%P96, HBL05, and A05, who are using slightly different disk
%conditions from one another.

In \S 3, we check our gas accretion prescription against P96 and HBL05, and
simulate the formation of a planet on a fixed orbit in an inviscid disk.
%with a fixed orbital radius of a planet, and compare our results with P96 and HBL05.
To make the comparison easier, we adopt
the same initial conditions as P96 and HBL05 --- the viscosity
parameter is \(\alpha=0\), the gas surface mass density profile is
\(\Sigma_{gas}=700(r/(5.2 \ AU))^{-2}=18950(r/AU)^{-2}\), the disk
temperature is \(T=150(r/5.2 AU)^{-1/2}=342(r/AU)^{-1/2}\), and the
disk extends from 0.25 to 50 AU. We use both a standard \(\kappa=1
\ {\rm cm^2 \ g^{-1}}\) and a reduced \(\kappa=0.03 \ {\rm cm^2 \
g^{-1}}\) opacity for our runs.

In \S 4, we include the effect of disk evolution and planet
migration, and compare our results with A05. Again, we choose the
same initial conditions as A05 to make the comparison easier, which
is the same as \S 3 except \(\alpha=2\times 10^{-3}\), and
\(\Sigma_{gas}=525(r/(5.2 \ AU))^{-2}=14196(r/AU)^{-2}\).
%In all of these cases, a disk extends from 0.25 to 50 AU, and the
%initial disk temperature is \(T=150(r/5.2
%AU)^{-1/2}=342(r/AU)^{-1/2}\).
%In \S 3, the disks are treated to be inviscid (i.e. the viscosity parameter is set
%to zero,) while in \S 4, the disks are assumed to have the viscosity of \(\alpha=2\times 10^{-3}\).

Finally in \S 5, we study the effect of a dead zone on planet
migration and formation. We use the same disk model as MPT07 ---
\(\Sigma_{gas}=\Sigma_0(r/AU)^{-3/2}\) with \(\Sigma_0=10^3\) or
\(10^4 \ {\rm g \ cm^{-2}}\), and the disk temperature model by
\cite{Robberto02}, which takes account of the effect of a cluster
environment. In this model, the disk extends from 0.02 to 100 AU.
%We set up the initial disks based on our
%previous work (MP06) as in MPT06, and study the effects of a dead
%zone on planet formation.
Throughout this section, we use the viscosity parameter
\(\alpha=10^{-5}\) inside the dead zones, and \(\alpha=10^{-2}\) for
the disk beyond it (active zones).
Here, the initial value of the outer dead zone radius is \(\sim 8.2\), and
\(15.8\) AU for \(\Sigma_0=10^3\), and \(10^4 \ {\rm g \ cm^{-2}}\), respectively.
%, which corresponds to assuming the initial value of the outer dead
%zone radius to be \(\sim 13\), and
%\(25\) AU respectively for \(\Sigma_0=10^3\) or \(10^4 \ {\rm g \
%cm^{-2}}\).
%Also, when the disks have no dead zones, the viscosity parameter is
%assumed to be constant throughout the disk (\(\alpha=10^{-2}\)).
These values of viscosity parameter agree well with the numerical
simulations done by \cite{Fleming03}.

The initial conditions of all runs are summarized in Table \ref{tb1}.
%
%%%%%%%%%%%%%%% Table 1
\begin{table*}
\center \caption{Initial conditions for all the runs. From left to
right columns, there are run name, initial gas surface mass density
($\Sigma_{\rm gas}=\Sigma_0 (a/AU)^{-p} \ [{\rm g \ cm^{-2}}]$),
disk temperature ($T=T_0 (a/AU)^{-q}$ [K]), planetesimal size,
viscous alpha values in active and dead zones, opacity, and torque
softening parameter. \label{tb1}}
\begin{tabular}{|c|c|c|c|c|c|c|}
\hline
Run &
$(\Sigma_0 \ [{\rm g \, cm^{-2}}], \ p)$ &
%of $\Sigma_{\rm gas}=\Sigma_0 (a/AU)^{-p} \ {\rm g \ cm^{-2}}$  &
$(T_0 \ [{\rm K}], \ q)$ &
%of $T=T_0 (a/AU)^{-q}$ &
planeteismal size &
$(\alpha_{\rm active}, \ \alpha_{\rm dead})$ &
$\kappa \ [{\rm cm^2 \, g^{-1}}]$ &
$B$
\tabularnewline
\hline
\hline
A &
$(10^3, \ 1.5)$ &
MPT07 &
N/A &
$(10^{-2}, \ 10^{-5})$ &
N/A &
N/A
\tabularnewline
\hline
B1 &
$(1.9 \times 10^4, \ 2)$ &
$(342, \ 0.5)$  &
100 m &
(0, N/A) &
$1$ &
N/A
\tabularnewline
\hline
B2 &
$(1.9 \times 10^4, \ 2)$ &
$(342, \ 0.5)$  &
1 km &
(0, N/A) &
$1$ &
N/A
\tabularnewline
\hline
B3 &
$(1.9 \times 10^4, \ 2)$ &
$(342, \ 0.5)$  &
10 km &
(0, N/A) &
$1$ &
N/A
\tabularnewline
\hline
B4 &
$(1.9 \times 10^4, \ 2)$ &
$(342, \ 0.5)$  &
100 km &
(0, N/A) &
$1$ &
N/A
\tabularnewline
\hline
C1 &
$(1.9 \times 10^4, \ 2)$ &
$(342, \ 0.5)$  &
100 m &
(0, N/A) &
$0.03$ &
N/A
\tabularnewline
\hline
C2 &
$(1.9 \times 10^4, \ 2)$ &
$(342, \ 0.5)$  &
1 km &
(0, N/A) &
$0.03$ &
N/A
\tabularnewline
\hline
C3 &
$(1.9 \times 10^4, \ 2)$ &
$(342, \ 0.5)$  &
10 km &
(0, N/A) &
$0.03$ &
N/A
\tabularnewline
\hline
C4 &
$(1.9 \times 10^4, \ 2)$ &
$(342, \ 0.5)$  &
100 km &
(0, N/A) &
$0.03$ &
N/A
\tabularnewline
\hline
D1 &
$(1.4 \times 10^4, \ 2)$ &
$(342, \ 0.5)$  &
10 km &
($2 \times 10^{-3}$, N/A) &
$1$ &
0.6
\tabularnewline
\hline
D2 &
$(1.4 \times 10^4, \ 2)$ &
$(342, \ 0.5)$  &
100 m &
($2 \times 10^{-3}$, N/A) &
$1$ &
0.6
\tabularnewline
\hline
D3 &
$(1.4 \times 10^4, \ 2)$ &
$(342, \ 0.5)$  &
100 m &
($2 \times 10^{-3}$, N/A) &
$1$ &
0.9
\tabularnewline
\hline
E1 &
$(10^3, \ 1.5)$ &
MPT07  &
100 m &
$(10^{-2}, \ 10^{-5})$ &
$1$ &
0.6
\tabularnewline
\hline
E2 &
$(10^3, \ 1.5)$ &
MPT07  &
100 m &
$(10^{-2}, \ 10^{-5})$ &
$0.03$ &
0.6
\tabularnewline
\hline
F1 &
$(10^4, \ 1.5)$ &
MPT07  &
100 m &
$(10^{-2}, \ 10^{-5})$ &
$1$ &
0.6
\tabularnewline
\hline
F2 &
$(10^4, \ 1.5)$ &
MPT07  &
100 m &
$(10^{-2}, \ 10^{-5})$ &
$0.03$ &
0.6
\tabularnewline
\hline
G1, \& G2 &
$(10^3, \ 1.5)$ &
MPT07  &
100 m &
$(10^{-2}, \ 10^{-5})$ &
$0.03$ &
0.6
\tabularnewline
\hline
%H1, \& H2 &
H1 &
$(10^4, \ 1.5)$ &
MPT07  &
100 m &
$(10^{-2}, \ 10^{-5})$ &
$0.03$ &
0.6
\tabularnewline
\hline
%H3 &
H2 &
$(10^4, \ 1.5)$ &
MPT07  &
100 m &
$(10^{-2}, \ 10^{-5})$ &
$1$ &
0.6
\tabularnewline
\hline
%\hline
\end{tabular}
\end{table*}
%%%%%%%%%%%%%%% Table 1 END
%
%--- Section 3 ------------------------------------------------------------------------------
%
\section{Planet Formation in non-evolving protoplanetary disks}
In this section, we check our gas accretion prescription against previous studies.
We simulate {\it in situ} planet formation in a
non-evolving protoplanetary disk by adopting the same initial conditions as P96 and HBL05,
and compare our results with theirs.
%We adopt the same initial conditions as these studies.
The disk models in this section assume an inviscid disk
(i.e. disk viscosity parameter is \(\alpha=0\) throughout the disk)
which extends from 0.25 to 50 AU (see \S 2.3 for details).
\subsection{Effect of the planetesimal size}
%Here, we reproduce the results of \cite{Pollack96}.
%
\begin{figure}
%\epsscale{2}\plottwo{figs4/plmass_p96_270_testKH.eps}{figs4/accr_rate_p96_270_testKH.eps}
\epsscale{2}\plottwo{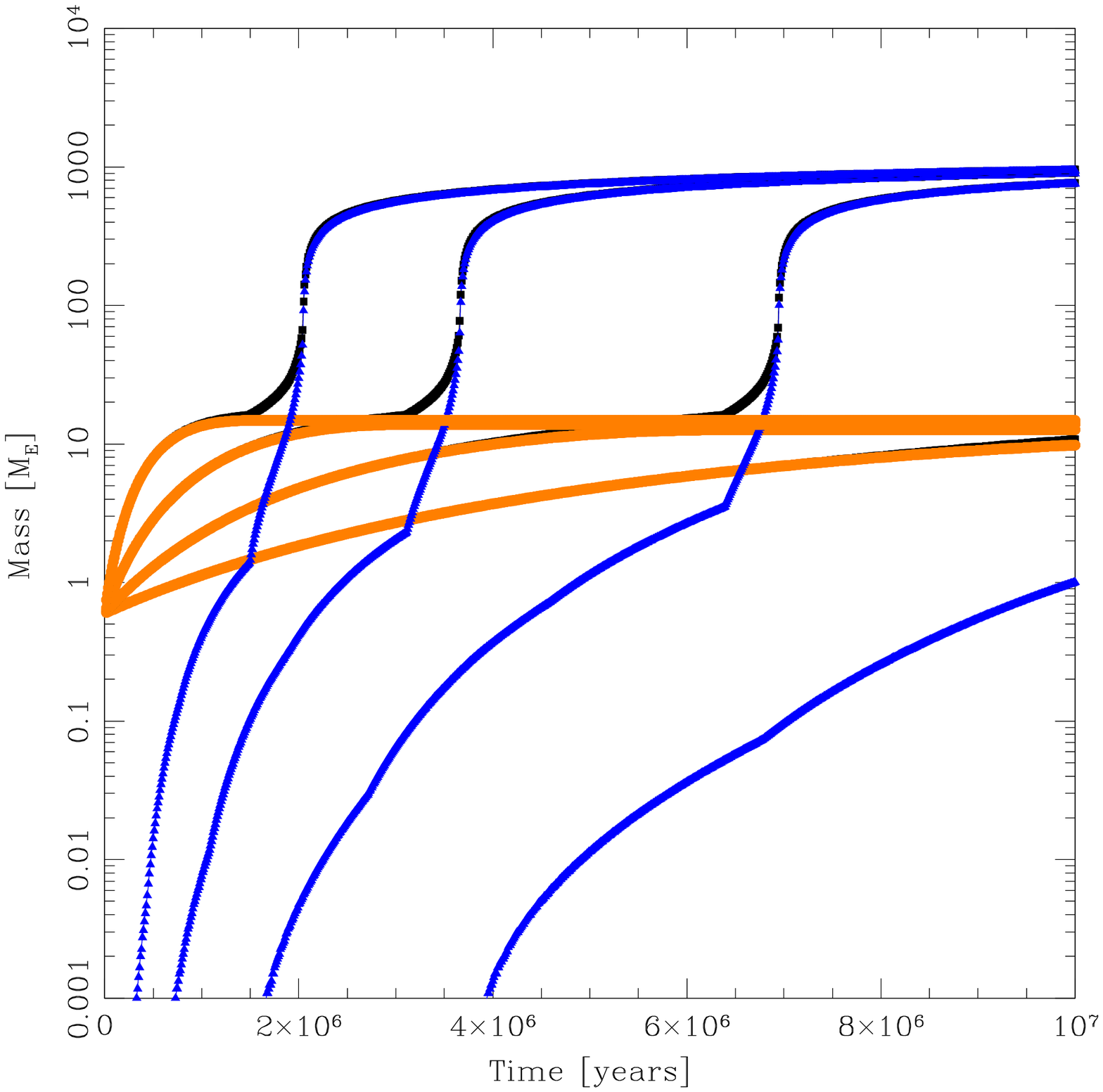}{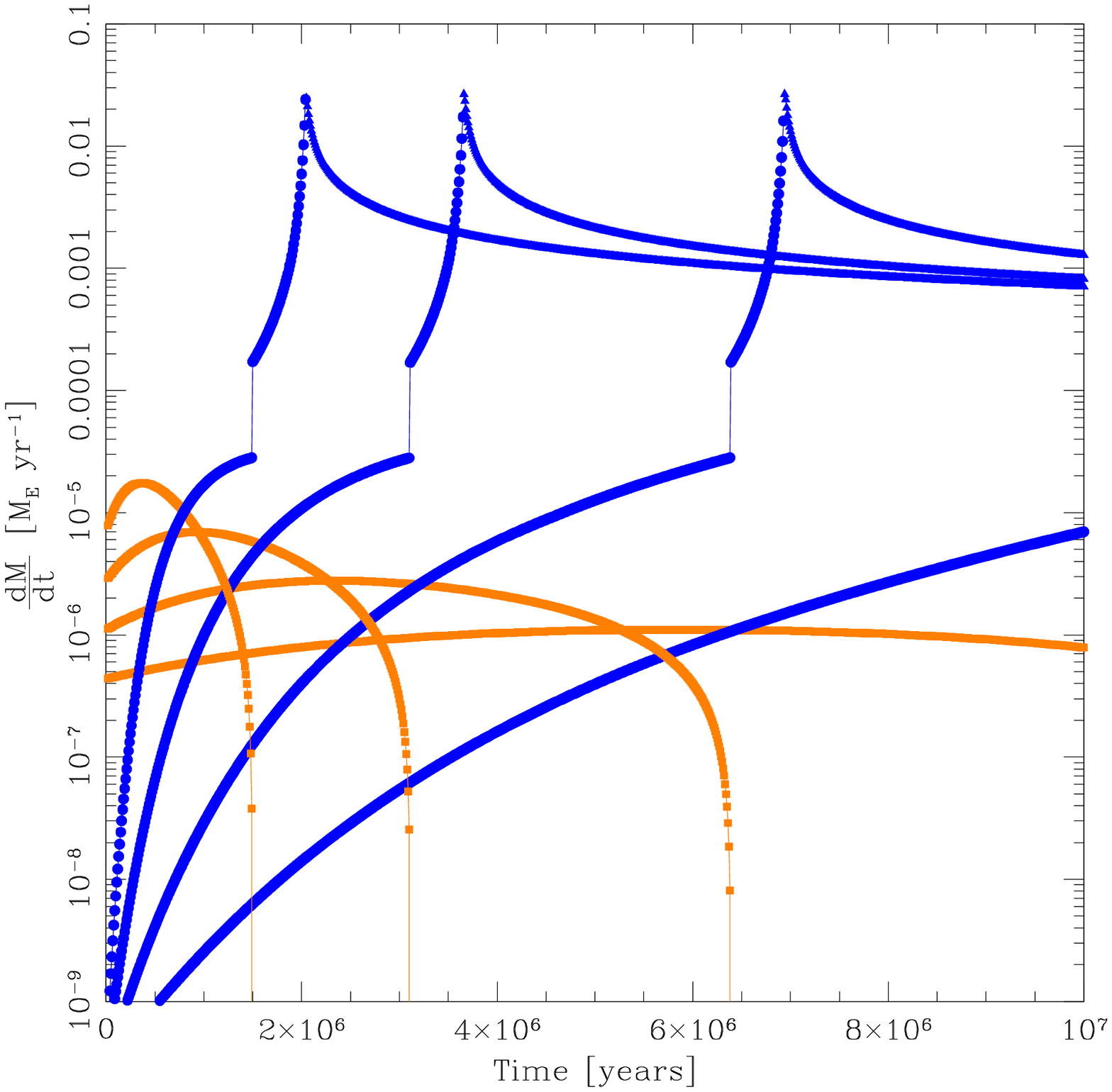}
%still play a fundament role in saving planetary systems.
%\includegraphics[width=7cm]{figs3/plmass_p96_270.eps}
%\includegraphics[width=7cm]{figs_chap6/test834_masstime.eps}
%%\includegraphics[width=7cm]{figs/test834_masstime.eps}
%\center
%\unitlength1cm
%\begin{minipage}[t]{15cm}
%%\hspace{0.5cm}
%\begin{minipage}[h]{15cm}
%\begin{picture}(1,6)
%\includegraphics[width=6.5cm]{figs4/plmass_p96_270_testKH.eps}
%\end{picture}
%\hspace{3.5cm}
%\begin{picture}(7,6)
%\includegraphics[width=6.5cm]{figs4/accr_rate_p96_270_testKH.eps}
%\end{picture}
%\end{minipage}
%\end{minipage}
%\begin{minipage}[t]{15cm}
%\vspace{0.5cm}
%
\caption[Mass evolution of a protoplanet at a fixed orbit]{Evolution of
planetary masses and accretion rates for planet at fixed disk orbital radius
of 5.2 AU (RunB1-4). Top: Evolution of planetary mass which is initially
\(0.6 M_E\). The planet has a fixed orbit at 5.2 AU from the central
star. The black curves show the total masses of planets, the orange
curves show the core masses, and the blue curves show the
envelope masses. Different sets of curves represent planetary growth
with different planetesimal sizes (\(100 \ {\rm m}, \ 1, \ 10, \
{\rm and} \ 100 \ {\rm km}\).)  The accretion time becomes shorter
for smaller planetesimals. The models with planetesimal size about
10 km produce a similar result to P96, and $10H\infty$ in HBL05.
Bottom: Corresponding mass accretion rates.
Orange curves are core accretion rates, and blue curves are gas
accretion rates from Eq. \ref{tau1}, \ref{tau2}, and \ref{tau3}.
%while blue curves are subdisk accretion rates from Eq. \ref{tau3}.
 \label{P96}}
%\end{minipage}
\end{figure}
%
%**** SOKO: I can't see any grey lines in this figure, nor the dashed
%Lines in top panel.  I only see solid black, orange and blue. Typo
%Here? ****

%First, we compare our gas accretion prescription with previous
%studies, and show the effect of the planetesimal size on planet
%formation timescale.
In this subsection, we focus on the effect of the planetesimal size
on planet formation timescale. Generally speaking, smaller
planetesimals lead to faster accretion because they are subject to
the stronger damping of random velocities, and hence have larger
cross-sections for accretion by a protoplanet.
%have the larger accretion cross section of a protoplanet.

P96 investigated the giant planet formation in
a non-evolving disk, and found that a Jupiter-like planet can form
{\it in situ} within about 8 Myr, while a similar study with
an improved equation of state and opacity tables by HBL05
predicts 6 Myr.
We use the same initial conditions as P96 and HBL05, and run four
different simulations by changing only the planetesimal size from
100 m to 100 km (RunB1-4).  RunB4 corresponds to the fiducial model
in P96, and $10H\infty$ in HBL05.
%study the difference in planet formation timescale depending
%on the different planetesimal sizes.

Fig. \ref{P96} shows the results of our simulations of gas and
planetesimal accretion by a protoplanetary core with \(0.6 M_E\) at
a {\it fixed orbit} (\(5.2\) AU). The top panel is the time evolution
of core, envelope, and total masses of a protoplanet for a
planetesimal size of 100 m, 1 km, 10 km, and 100 km from left to
right.  The bottom panel shows the time evolution of core and envelope
mass accretion rates for the corresponding runs. Once the core
accretion ceases, the transition from Eq. \ref{tau1} to \ref{tau2}
occurs, and the envelope accretion rate increases, which is marked
by a jump. Planets approach their final masses as the gas
accretion slows down due to the subdisk accretion and a gap-opening.
%
%We choose the initial conditions so that we could reproduce the
%results of P96.
%As described in the previous section, planet formation via core
%accretion is composed of the core building phase and the gas
%accretion phase.
%By comparing the left and bottom panels, we find that the gas accretion rate becomes comparable to the core accretion rate
%when the core mass reaches \(\sim 10 M_E\).

As the previous studies have shown, we find that the core building
takes longer for a swarm of larger planetesimals, and so does the
time until the start of the rapid gas accretion. As 
it was mentioned earlier, this is because smaller planetesimals have larger 
accretion cross-sections by a protoplanet. For 100 m-size
planetesimals, a Jupiter-like planet forms within about 2 Myr, while
for 10 km-size planetesimals, it takes about 7 Myr, which is roughly
the same as the time estimated for the fiducial model of P96 (\(\sim
8\) Myr) or the corresponding model ($10H\infty$) in HBL05 (\(\sim
6\) Myr).
%Smaller planetesimals lead to the faster accretion because
%they are subject to the stronger damping of random velocities, and
%hence have the larger accretion cross section of a protoplanet.

Now we compare our results with the fiducial model of P96 and
$10H\infty$ in HBL05. P96(HBL05) obtained the estimated formation
timescale of 8(6) Myr for 100 km-size planetesimals.   In our run,
the corresponding case (RunB4) did not achieve a Jupiter mass within the
simulation time (10 Myr). Moreover, although the mass evolution
profiles look qualitatively similar to the ones in P96 or HBL05, the
details appear to be different. Phase 1 and 2 in their cases are
merged in our case.  The difference is apparent by comparing the
bottom panel of Fig. \ref{P96} with Fig. 1b in P96. In their case,
rapid planetesimal accretion (Phase 1) ceases when a protoplanet
depletes its planetesimal feeding zone, and this is followed by slow
gas accretion (Phase 2), which lasts until the planet obtains the
crossover mass for rapid gas accretion (Phase 3).
Since they assume rapid planetesimal accretion,
%where the relative velocity of the planetesimals is smaller than the escape speed from
%the surface of the protoplanet (\(v_{rel}<v_{esc}\)),
the planetesimal accretion rate is very high during Phase 1 (\(dM/dt
\sim 10^{-4}-10^{-2}  \ {\rm M_E/yr} \)), and drops to \(\sim
10^{-6} \ {\rm M_E yr^{-1}}\) during Phase 2 due to planetesimal
depletion in the feeding zone (see Fig. 1b of P96.) Their
protoplanetary core grows from $0.6 M_E$ to $\sim 10 M_E$ during Phase
1 (within 0.5 Myr).

However, as we mention in \S 2, protoplanets more massive than a few
times $10^{-5} M_E$ are expected to grow {\it oligarchically}. In
our simulations, since we assume slower, oligarchic growth from the
start, the planetesimal accretion rate is low all the time \(\sim
10^{-5}-10^{-6} \ {\rm M_E yr^{-1}}\) and drops to zero once the
planetesimals are depleted. As a result, the core mass does not
reach $\sim 10 M_E$ for a few Myr in RunB3. The difference in core
growth prescription may not seem to have a significant effect on a
non-migrating planet in the inviscid disk, but becomes important for
a migrating planet in an evolving disk. We discuss this further in
\S 4 and 5.
%
%A jump in gas accretion rate occurs
%because solid accretion ends (cf. Eq. \ref{tau1} and \ref{tau2}.)
%The planetary masses approach a final mass when the subdisk
%accretion starts and gas accretion slows down.

\subsection{Effect of the opacity}
A reduced opacity in the gaseous envelope of a protoplanet can lead
to the faster gas accretion (P96 and INE00). HBL05
extended the work of P96, and studied the effects of disk's opacity
as well as a planetary core mass. By assuming \(2\%\) of the opacity
of interstellar grains, they found that the gas accretion time
becomes significantly shorter (\(1-4.5 \ {\rm Myr}\)).  This effect
is reflected in the gas accretion timescale we adopt (see Eq.
\ref{tau1} and \ref{tau2}).

%Our result can be directly compared with the model of \(10L\infty\)
%in HBL05 for planetesimal size of \(100\) km.
%They assumed the planetesimal surface mass density is
%\(\Sigma_{solid}=10 \ (r/(5.2AU))^{-2}=270.4 \ (r/AU)^{-2}\) with
%planetesimal size of \(100\) km.
%and that the planetesimal accretion is cut off at \(10 M_E\).
We keep the definition of our solid surface mass density, and change
the size of planetesimals from 100 m to 100 km as in the last
subsection (RunC1-4).
RunC4 corresponds to the model of \(10L\infty\) in HBL05.
%stop the planetesimal accretion at \(10 M_E\).
Here, we adopt a reduced, constant opacity of \(\kappa=0.03 \ {\rm
cm^{2} \ g^{-1}}\), which is an average opacity value obtained for
the model $10L\infty$ in HBL05 over a relevant range of temperature
of protoplanetary disks (roughly \(100-1000\) K).
%The run with 100 km-size planetesimals
%RunC4 corresponds to the model of \(10L\infty\) in HBL05.

\begin{figure}
%\epsscale{2}\plottwo{figs4/plmass_h05_270_testKH.eps}{figs4/accr_rate_h05_270_testKH.eps}
\epsscale{2}\plottwo{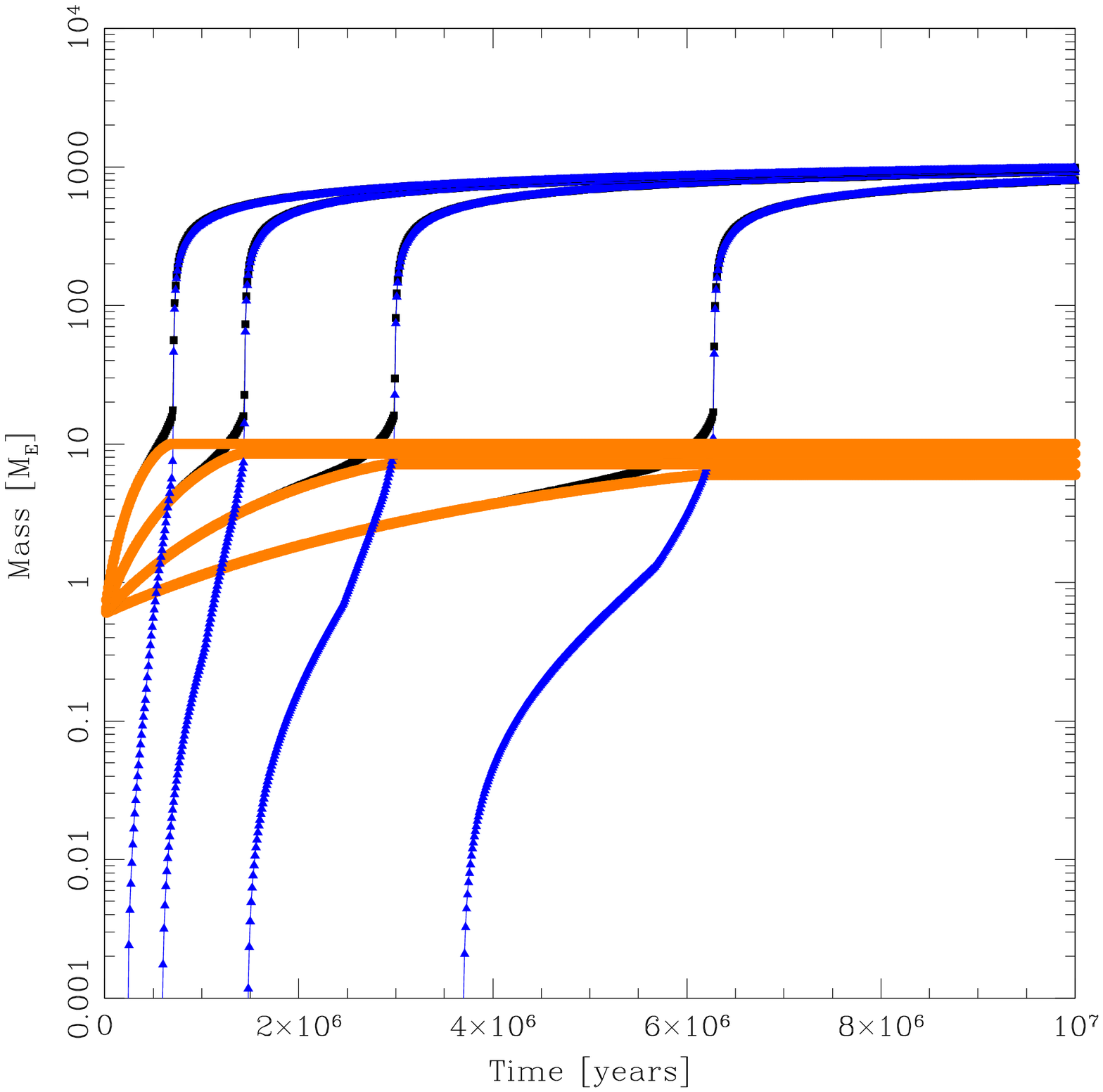}{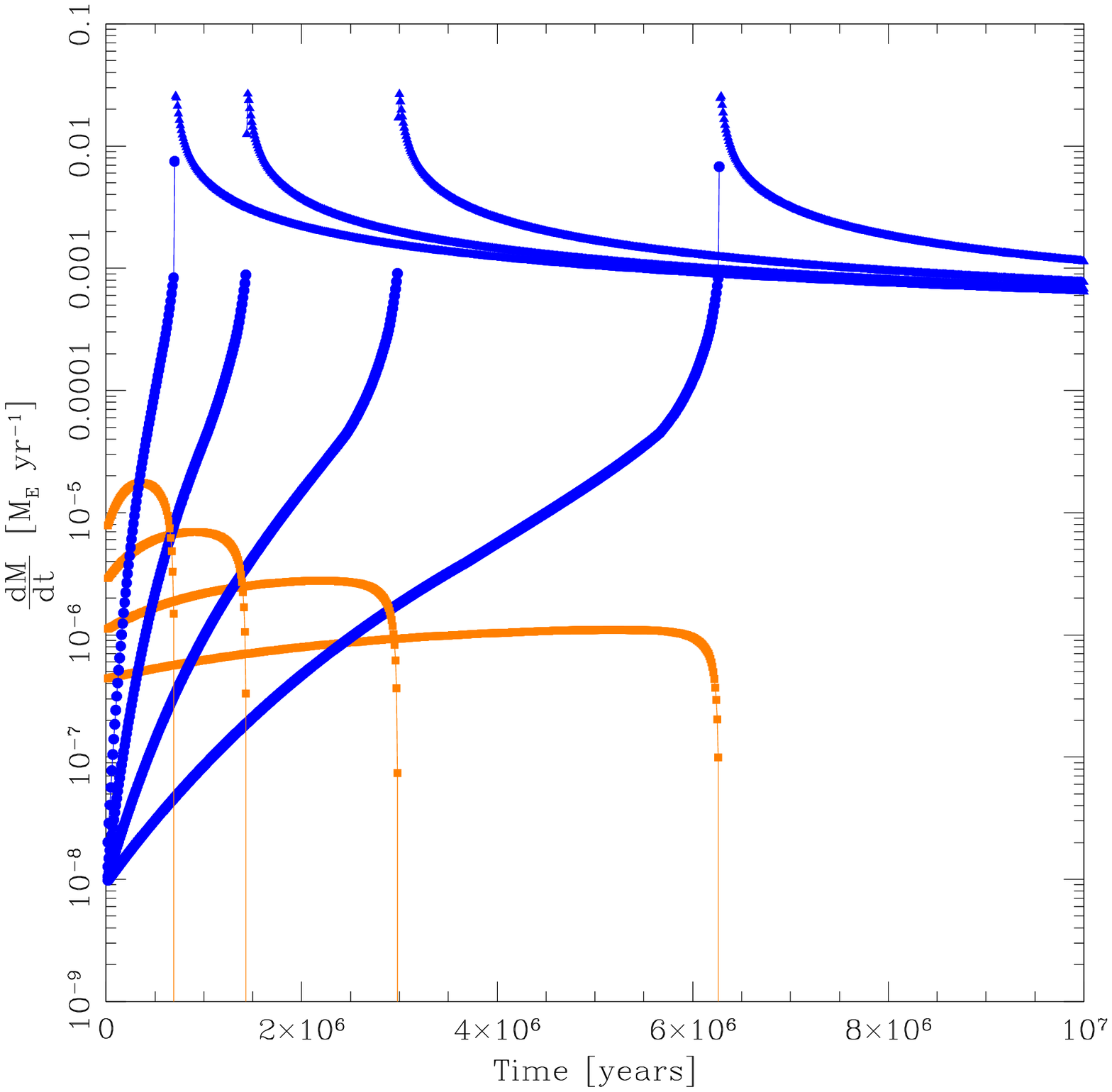}
%\includegraphics[width=7cm]{figs3/plmass_h05_270.eps}
%\includegraphics[width=7cm]{figs_chap6/test833correct_masstime.eps}
%%\includegraphics[width=7cm]{figs/test838_masstime.eps}
%
%\unitlength1cm
%\begin{minipage}[t]{9cm}
%%\hspace{0.5cm}
%\begin{minipage}[h]{9cm}
%\begin{picture}(1,6)
%\includegraphics[width=4.5cm]{figs4/plmass_h05_270_testKH.eps}
%\end{picture}
%\hspace{3.5cm}
%\begin{picture}(7,6)
%\includegraphics[width=4.5cm]{figs4/accr_rate_h05_270_testKH.eps}
%\end{picture}
%\end{minipage}
%\end{minipage}
%\begin{minipage}[t]{9cm}
%\vspace{0.5cm}
%
\caption[Mass evolution of a protoplanet at a fixed orbit in a disk
with a reduced opacity]{Effect of reduced opaciy on evolution
of planetary masses and accretion rates for planet at
fixed orbital radius of 5.2 AU (RunC1-4). Top: The same as Fig.
\ref{P96}, but with a disk opacity of \(\kappa=0.03 \ {\rm cm^2 \
g^{-1}}\). The model with planetesimal size about \(10\) km results
in a similar outcome to the model \(10L\infty\) in HBL05.
Bottom: Corresponding mass accretion rates.
% and a core growth stops at \(10 M_E\).
%This corresponds to the model \(10L10\) in \cite{Hubickyj05}.
\label{P96opa}}
%\end{minipage}
\end{figure}
The results of our simulations are shown in Fig. \ref{P96opa}.
Comparing Fig. \ref{P96} with \ref{P96opa}, we confirm that the
lower opacity results in the faster gas accretion. HBL05 showed that
the formation time could be as short as 2 Myr in a disk with 100 km
planetesimals with a reduction in opacity (see their Fig. 1). In our
simulations, a similar time scale to their model is obtained for a 10 km-size
planetesimal disk (RunC3), where formation takes about 3 Myr.
%, again, produces a good agreement with the
%HBL05 model, and shows that a planet starts a rapid gas accretion
%within about \(3 \times 10^6\) years.
This is more than twice faster than our corresponding run in \S 3.1 
with the standard opacity (RunB3). The formation time scales for our
10 km planetesimal disk models are about 1 Myr longer than those of
corresponding models in HBL05 with a 100 km-size planetesimal disk
($10L\infty$ and $10H\infty$). Thus, our gas accretion prescription agrees 
fairly well with previous studies like P96 and HBL05.
%
%--- Section 4 ------------------------------------------------------------------------------
%
%
\section{Planet formation in an evolving disk}
We now add disk evolution and planet migration to the formation
model described in the previous section, and study their effects on
planet formation to compare the results with A05.
%compare our results with \cite{Alibert05}.
In this section of the paper, the disk viscosity parameter is
assumed to be constant (\(\alpha=2\times 10^{-3}\)) throughout the
disk. Therefore, different from the previous section, a protoplanetary
disk accretes toward the central star.  Note that this model does
not include a dead zone.  The other initial conditions of the disk
models are explained in \S 2.3, and summarized in Table \ref{tb1}.
%
%Similar to the last section, we use the model of a disk stretching
%from \(0.25\) to \(50\) AU which contains the
%%\(10\) km size
%planetesimals with the surface mass density of
%\(\Sigma_{solid}=270(r/AU)^{-2}\).
%
%To compare our results with the one in \cite{Alibert05}, we choose
%the gas surface mass density of  \(\Sigma_{gas}=525(r/(5.2
%AU))^{-2}=14196(r/AU)^{-2}\).

A05 adopted the migration speed obtained by \cite{Tanaka02}, and
reduced the migration rate artificially by a factor of \(0.1-0.01\).
To incorporate the idea, but to keep the generality, we use the
torque expression presented by \cite{Menou04}, and adjust the
softening parameter \(B\), which mimics the torque reduction effect
with a vertical height (also see Appendix B in MPT07).

\begin{figure}
%\includegraphics[width=7cm]{figs_chap6/test836correct.eps}
%\includegraphics[width=7cm]{figs3/a05_270_B06_100m_8au.eps}
%\epsscale{2}\plottwo{figs4/disk_a05_270_B06_10km.eps}{figs4/plmass_diskmass_diskaccr_a05_270_B06_10km.eps}
\epsscale{2}\plottwo{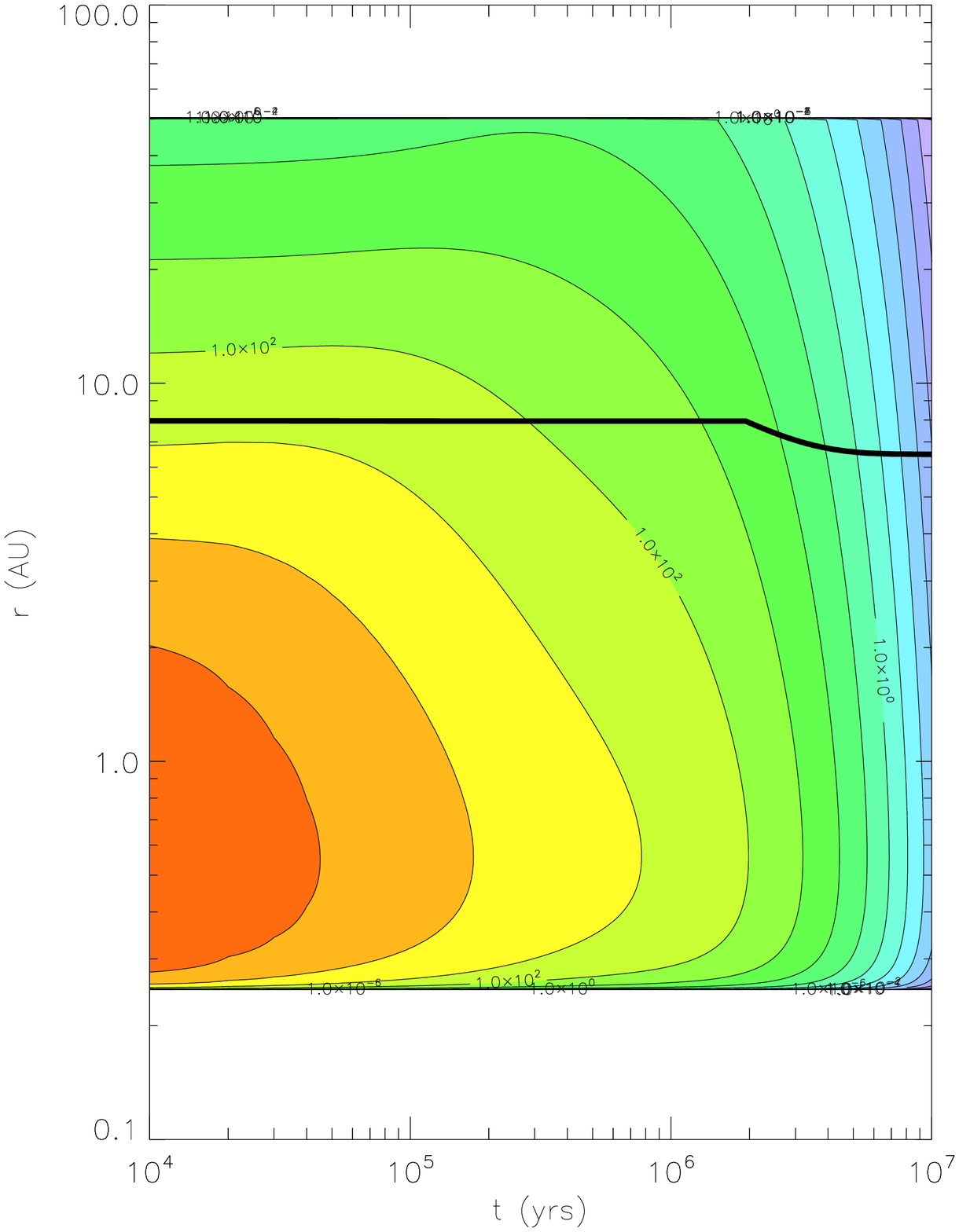}{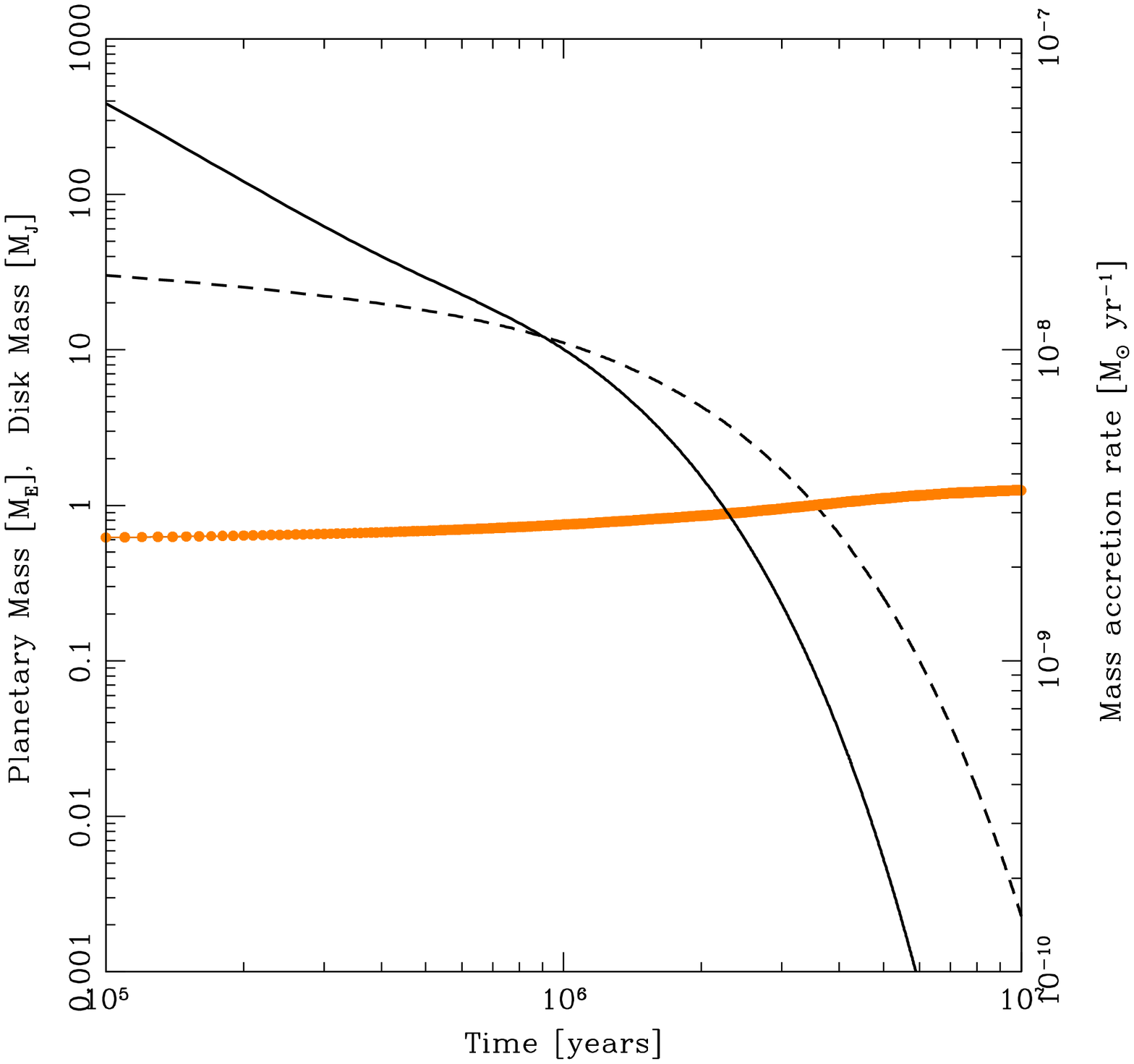}
%\plottwo{fig6a.eps}{fig6b.eps}
%
%\unitlength1cm
%\begin{minipage}[t]{9cm}
%%\hspace{0.5cm}
%\begin{minipage}[h]{9cm}
%\begin{picture}(1,6)
%\includegraphics[width=4.5cm]{figs4/disk_a05_270_B06_10km.eps}
%\end{picture}
%\hspace{3.5cm}
%\begin{picture}(7,6)
%\includegraphics[width=4.5cm]{figs4/plmass_diskmass_diskaccr_a05_270_B06_10km.eps}
%\end{picture}
%\end{minipage}
%\end{minipage}
%\begin{minipage}[t]{9cm}
%%\vspace{0.5cm}
%
\caption[Evolution of a disk with a migrating protoplanet
(B=0.6)]{Evolution of planet and disk for planetesimal sizes of 10 km
and no DZ (RunD1). Top: Evolution of a planet with an
initial mass of \(0.6 M_E\) in an evolving disk with no dead zone.
The disk viscosity parameter is set to \(\alpha=2\times 10^{-3}\),
and the planetesimal size is 10 km. We use our fiducial softening
parameter \(B=0.6\). There is very little growth of a planet, and as
a result, the planet does not migrate much.
%The protoplanet plunges into the star within \(2\times 10^6\) years,
%and its final mass is roughly \(20 \ M_E\).
Bottom: Corresponding evolution of planet and disk masses as well as
disk accretion rate onto the star. Solid, and dashed curve shows
disk accretion rate, and disk mass, respectively. Thick orange, and
black curves are on top of each other, and show core, and total
planetary mass, respectively. \label{fig6}}
%\end{minipage}
\end{figure}
First, we show our fiducial migration case with \(B=0.6\), which
roughly corresponds to the migration speed estimated by
\cite{Tanaka02}.
%
%First, we show the case of a planet growth without slowing down the
%migration.
Fig. \ref{fig6} shows the growth of a planet with an initial mass of
\(0.6 M_E\) in a 10 km-size planetesimal disk (RunD1). The
planetesimal accretion is very slow as it can be seen in the bottom
panel, and therefore the protoplanet does not migrate much. The
final mass is about \(1 \ M_E\), only barely increased from \(0.6 \
M_E\). Also plotted in the bottom panel of Fig. \ref{fig6} are the
disk mass evolution (dashed line) and disk accretion rate onto the
central star (solid thin line). It is interesting to compare this
figure with the disk evolution with a dead zone (see the bottom panel
of Fig. \ref{fig3}). Viscous disk accretion rate, and hence disk
mass, decrease gradually rather than being nearly constant over a
long time as in the case with a dead zone. If all disks have high
alpha values $\alpha>10^{-3}$, very massive disks may be necessary
to explain the existence of long-lived, strong accretors.
%In such an evolution
%scenario, photoevaporation may be necessary to explain the observed
%rapid dispersal of a gas disk, as briefly discussed in \S 2.2.

%The planet migrates inward and plunges into the central star within about 2 Myr.
%Although the planetary mass increases rapidly
%as the protoplanet migrates, the final mass is about \(15 \ M_E\),
%and there is little gas accretion. Although the final gas accretion
%rate becomes comparable to the solid accretion rate (\(\sim 5 \times
%10^{-5} \ {\rm M_E \ yr^{-1}}\)), migration is too fast in this case
%to make a Jupiter-like planet.
%Since we have not slowed the
%migration artificially as in \cite{Alibert05}, the planet starts
%migrating from \(8\) AU and plunges into the central star within
%\(\sim 2\times 10^5\) years ({\bf use new figure \& value!}).
%
\begin{figure}
%\includegraphics[width=7cm]{figs_chap6/test836correct_masstime.eps}
%\includegraphics[width=7cm]{figs3/a05_270_B06_10km_8au.eps}
%\epsscale{2}\plottwo{figs4/disk_a05_270_B06_100m.eps}{figs4/plmass_diskmass_diskaccr_a05_270_B06_100m.eps}
\epsscale{2}\plottwo{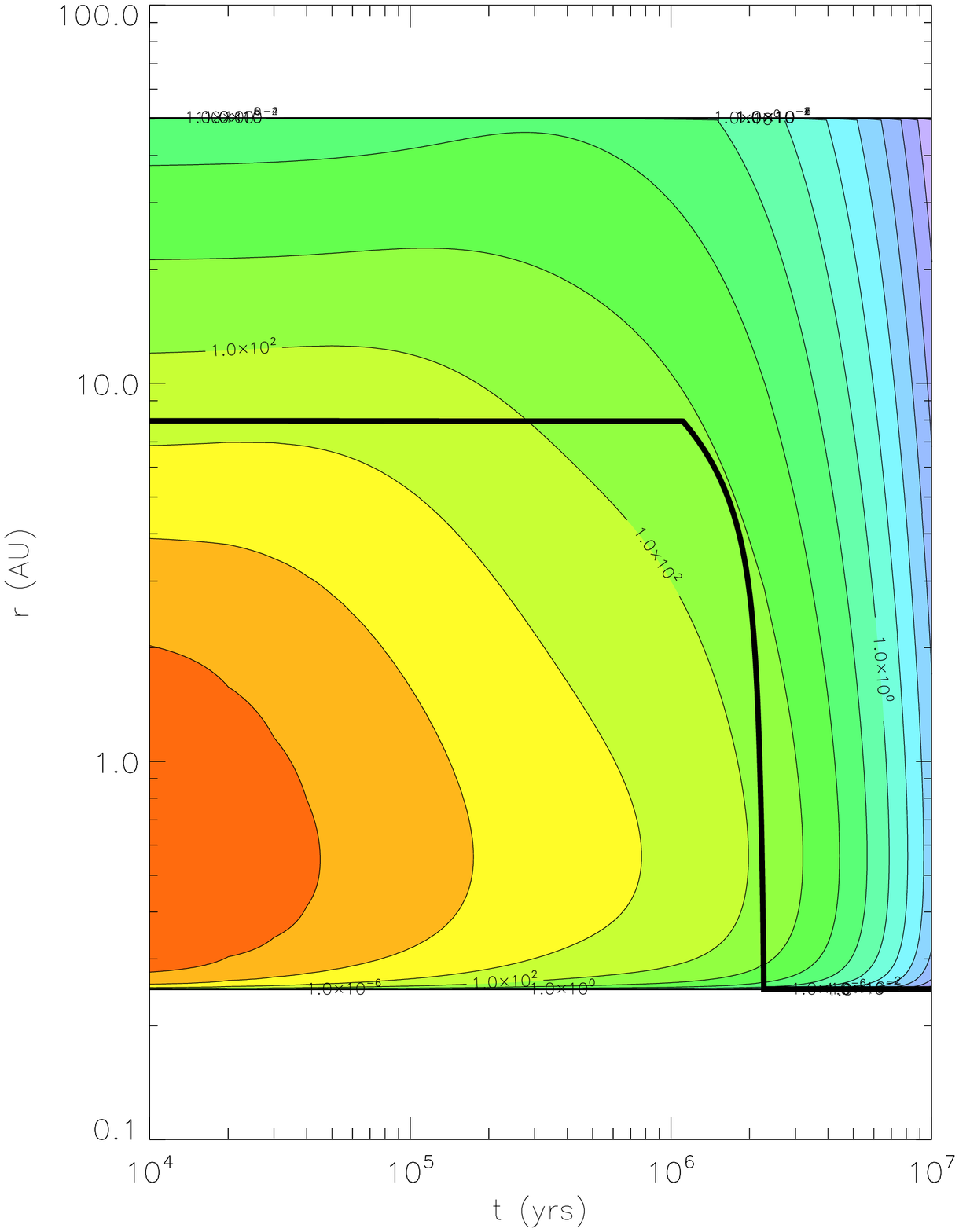}{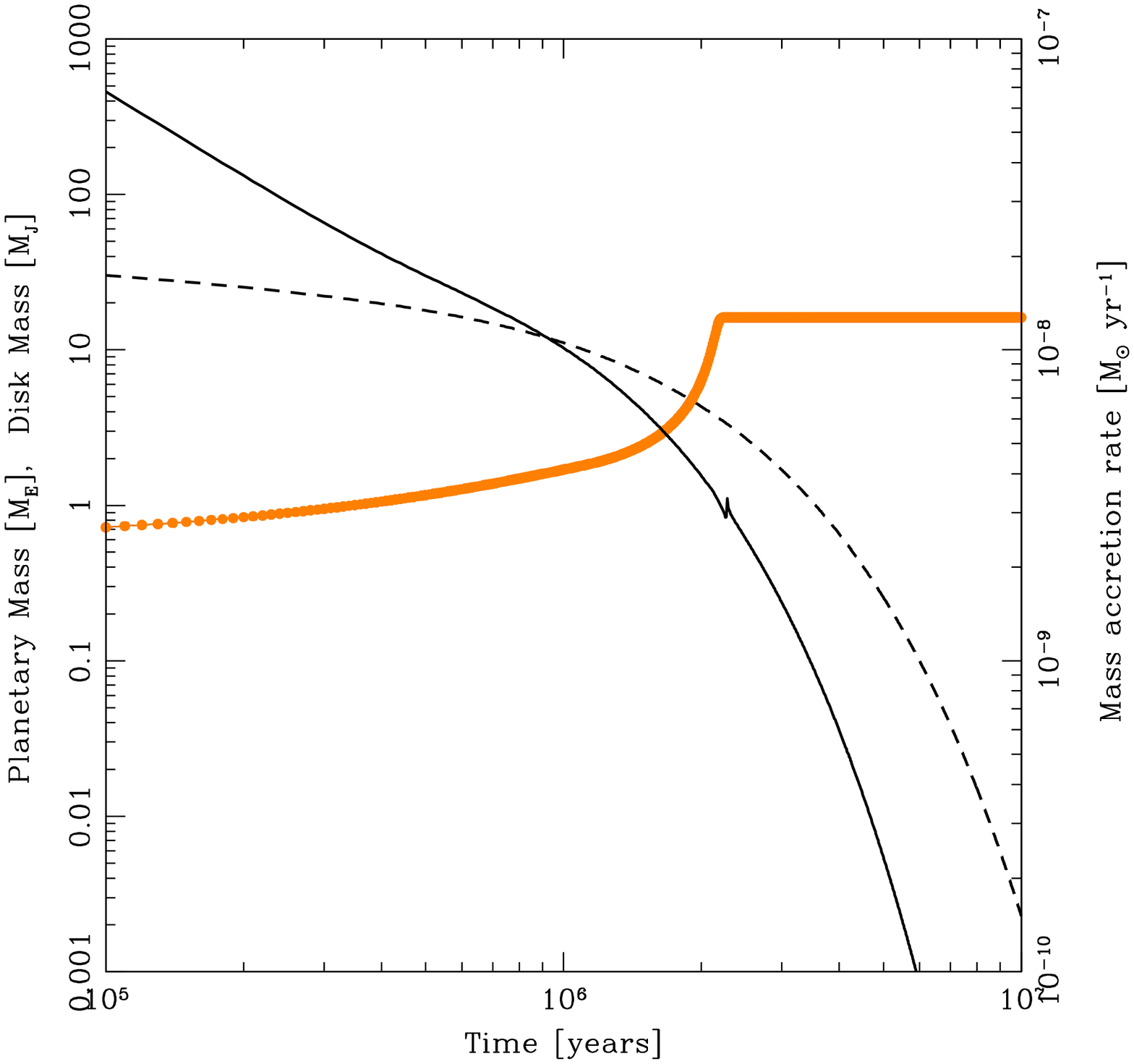}
%
%\unitlength1cm
%\begin{minipage}[t]{9cm}
%%\hspace{0.5cm}
%\begin{minipage}[h]{9cm}
%\begin{picture}(1,6)
%\includegraphics[width=4.5cm]{figs4/disk_a05_270_B06_100m.eps}
%\end{picture}
%\hspace{3.5cm}
%\begin{picture}(7,6)
%\includegraphics[width=4.5cm]{figs4/plmass_diskmass_diskaccr_a05_270_B06_100m.eps}
%\end{picture}
%\end{minipage}
%\end{minipage}
%\begin{minipage}[t]{9cm}
%%\vspace{0.5cm}
%
\caption[Mass evolution of a migrating protoplanet (B=0.6)]{Evolution of
planet and disk with no DZ (as in Fig. 6), except
with planetesimal sizes of 100 m (RunD2).
%Same as Fig. \ref{fig6}, but with the planetesimal size of 100 m.
Top: The protoplanet plunges into the star within 2 Myr,
and its final mass is roughly \(20 \ M_E\). Bottom:
Corresponding evolution of planet and disk masses as well as the disk
accretion rate onto the star. \label{fig7}}
%\end{minipage}
\end{figure}

Fig. \ref{fig7} shows the same run as Fig. \ref{fig6}, but with the
planetesimal size of 100 m (RunD2). The planet migrates inward, and
plunges into the central star within about 2 Myr. Although the
planetary mass increases rapidly as the protoplanet migrates, the
final mass is about \(15 \ M_E\), and there is little gas accretion.
Although the final gas accretion rate becomes comparable to the
solid accretion rate (\(\sim 5 \times 10^{-5} \ {\rm M_E \
yr^{-1}}\)), migration is too fast in this case to make a
Jupiter-like planet.
It should be noted, however, that the resolution of our simulation is good
down to ~0.1 AU, and that we don't define the inner dead zone edge.
Therefore, any planets which appear to be ``lost'' in our
simulations may still be alive if we improve our resolution, take
account of the inner dead zone edge, or include the star-planet
tidal interaction.
%
%
%The final planetary mass is about \(20 \ M_E\), and there is little gas accretion.
%Here, we don't see
%a characteristic plateau feature seen in Fig. \ref{P96}.  This is
%because the amount of mass within a feeding zone increases as the
%planet migrates inward.  In fact, the planetary mass reaches one
%third of the Jupiter mass just before it is swallowed by the central
%star.

\begin{figure}
%\includegraphics[width=7cm]{figs_chap6/test836correct_masstime.eps}
%\includegraphics[width=7cm]{figs/test836correct_masstime.eps}
%\epsscale{2}\plottwo{figs4/disk_a05_270_B09_100m.eps}{figs4/plmass_diskmass_diskaccr_a05_270_B09_100m.eps}
\epsscale{2}\plottwo{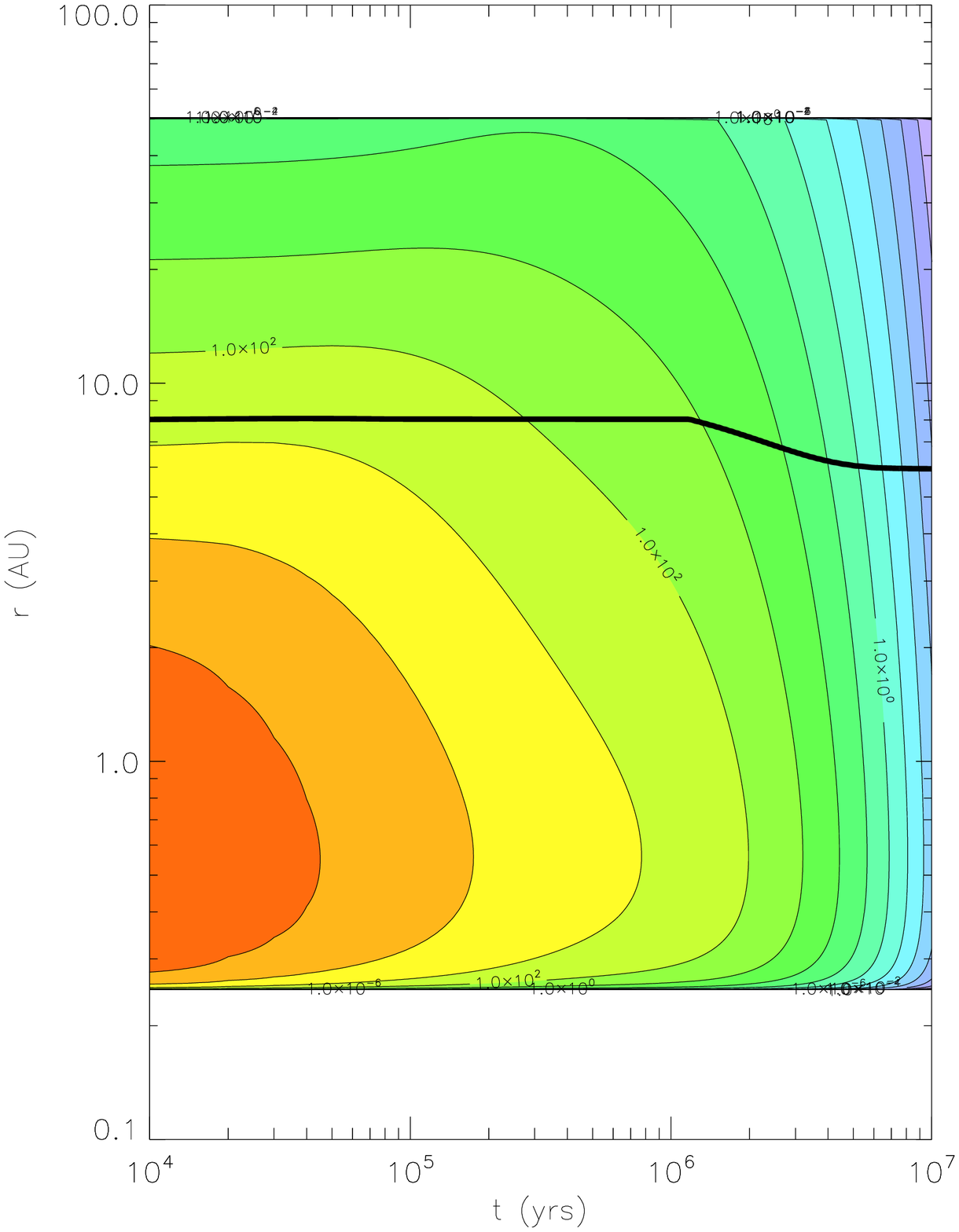}{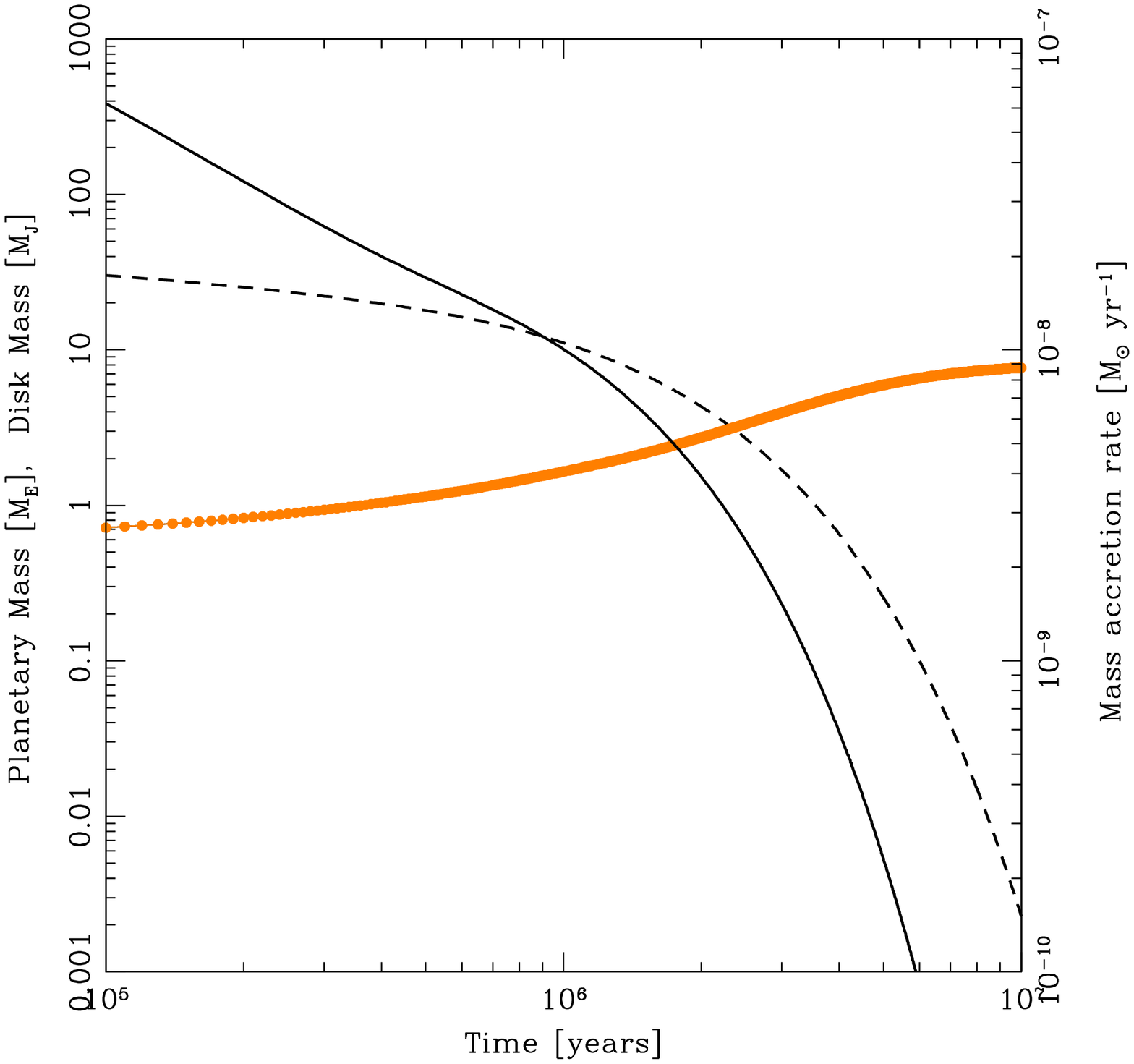}
%
%\unitlength1cm
%\begin{minipage}[t]{9cm}
%%\hspace{0.5cm}
%\begin{minipage}[h]{9cm}
%\begin{picture}(1,6)
%\includegraphics[width=4.5cm]{figs4/disk_a05_270_B09_100m.eps}
%\end{picture}
%\hspace{3.5cm}
%\begin{picture}(7,6)
%\includegraphics[width=4.5cm]{figs4/plmass_diskmass_diskaccr_a05_270_B09_100m.eps}
%\end{picture}
%\end{minipage}
%\end{minipage}
%
%\begin{minipage}[t]{9cm}
%%\vspace{0.5cm}
\caption[Mass evolution of a migrating protoplanet (B=0.9)]{Evolution of
planet and disk for planetesimal sizes of 100 m
and no DZ, but with larger softening parameter $B=0.9$ (RunD3).
%Evolution of a planet's mass which is initially \(0.6
%M_E\). With the softening parameter \(B=0.9\),
The planet migrates much slower, but grows only up to $9 \ M_E$.  \label{fig8}}
%\end{minipage}
\end{figure}
\begin{figure}
%\plotone{figs4/accr_rate_a05_270_B09.eps}
\plotone{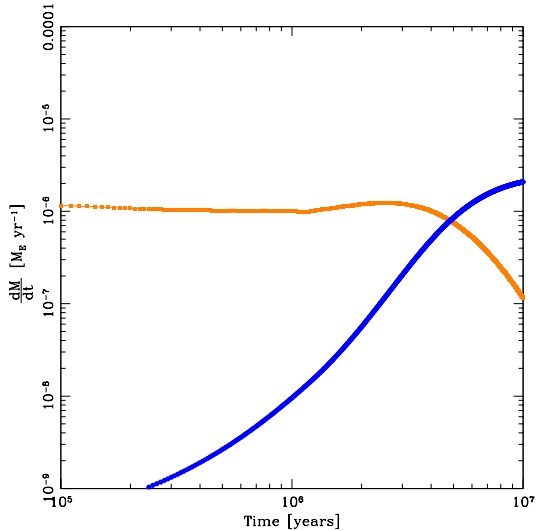}
\caption{Corresponding
mass accretion rate for Fig. \ref{fig8}.  Orange curve is the core
accretion rate, and blue curve is the envelope accretion rate.  As
the planetary mass grows, gas accretion rate increases and the two
accretion rates become comparable to each other. \label{fig88}}
\end{figure}
Now we show the case of slower migration by adopting a larger
softening parameter of \(B=0.9\) (RunD3), which roughly reproduces the
migration profile of A05 (see their \S 3.1.) Fig. \ref{fig8} shows such an
evolution in a disk with 100 m-size planetesimals. The planet
migrates more slowly, but grows only up to \(\sim 10 M_E\).  From disk
mass evolution (dotted line on the bottom panel), we can see that
formation of a Jupiter mass planet is impossible after 4 Myr, since
the disk mass falls below $M_J$.  In our run, the planet's mass is
$\sim 7 M_E$ at 4 Myr.
%
%Now we show the case similar to \cite{Alibert05}, where we
%artificially slow down the migration by a factor of 0.1.  Fig.
%\ref{Alibert05tr2} shows the growth of a planet in an evolving disk.
%The planet migration is slowed down, but the planet still plunges
%into the central star after \(\sim 1.6 \times 10^6\) years.
%
%Fig. \ref{Alibert052} shows the mass evolution of the same run.
%Again, the planetary mass increases rapidly as it migrates.  We see
%a characteristic plateau feature in this run, but it's much shorter
%compared to Fig. \ref{P96}.  Again, the planet acquires a mass about
%one third of Jupiter by the time it is swallowed by the central
%star.

This is very different from what A05 obtained, where the planet
starts opening a gap after \(0.8\) Myr at around 6 AU, and stops its
migration at around 5.5 AU.  This results from the difference in
solid accretion prescriptions as described briefly in the previous
section.  A05 followed P96 and adopted rapid planetesimal accretion
for the growth of the planetary core with an initial mass of $0.6
M_E$. Therefore, their core mass reaches several $M_E$ by 0.8 Myr,
while our core mass is about $2 M_E$ at that time.

In P96, the planetesimal accretion slows when the protoplanet depletes
the planetesimal feeding zone.  Therefore, during Phase 2,
their planetary mass increases at the gas accretion rate,
which is much smaller than the planetesimal accretion rate in Phase 1.
As a result, the planet formation timescale in P96 was uncomfortably long.
A05 overcame this problem by constantly
replenishing the planetary feeding zone due to migration, and
therefore increasing the total mass of a planet at the rapid
planetesimal accretion rate rather than the gas
accretion rate.  Thus, the planet accretes gas much quicker
in their case.
On the other hand, Fig. \ref{fig88} shows that the planetesimal
accretion rate in our model is a few orders of magnitude lower than
that during Phase 1 in P96, and comparable to that during Phase 2 in P96.
Gas accretion rate is even smaller, or at most comparable to
planetesimal accretion rate throughout the simulation.
Therefore, in our case, a protoplanet increases its mass on nearly
constant time scale all the time,
independent of whether it is migrating or not.
%is much lower in our case, and roughly comparable to
%the gas accretion rate, so that the growth rate of the total planetary mass is
%comparable to the gas/planetesimal accretion rate.
In short, for the {\it oligarchic} core accretion process, we find that the replenishment of the
planetesimal feeding zone due to migration does not help planetary
growth as much as it was shown by A05.
\section{Planet Formation in an evolving disk with a dead zone}
In this section, we combine all of the elements of protoplanetary
accretion and disk evolution presented above with the dead zone.
\subsection{Without planet migration}
%and see how its presence affects the planet formation scenario.
%
\begin{figure}
%\epsscale{2}\plottwo{figs4/plmass_diskmass_diskaccr_5au_opa1_sigma1e3.eps}{figs4/plmass_diskmass_diskaccr_5au_opa003_sigma1e3.eps}
\epsscale{2}\plottwo{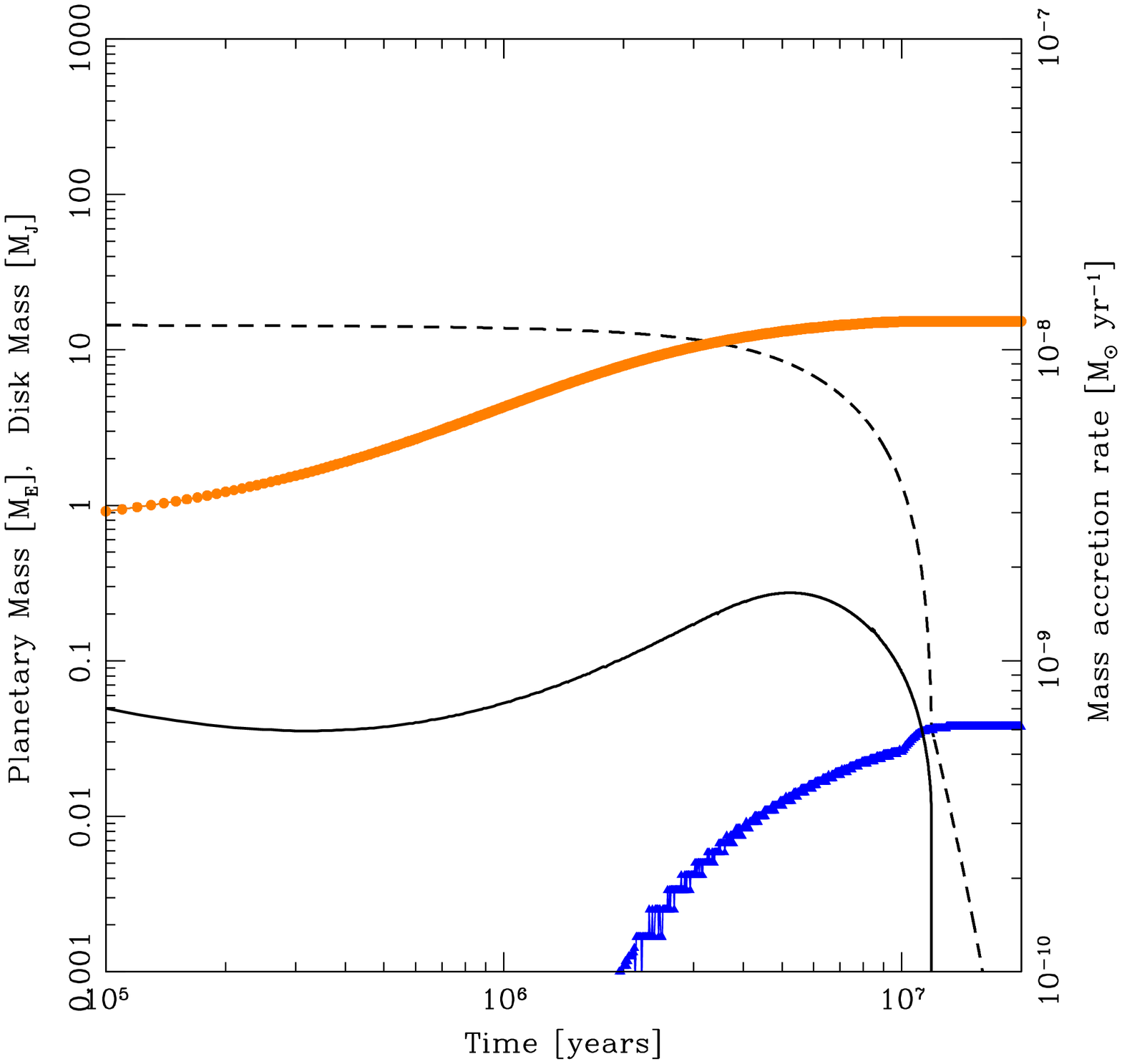}{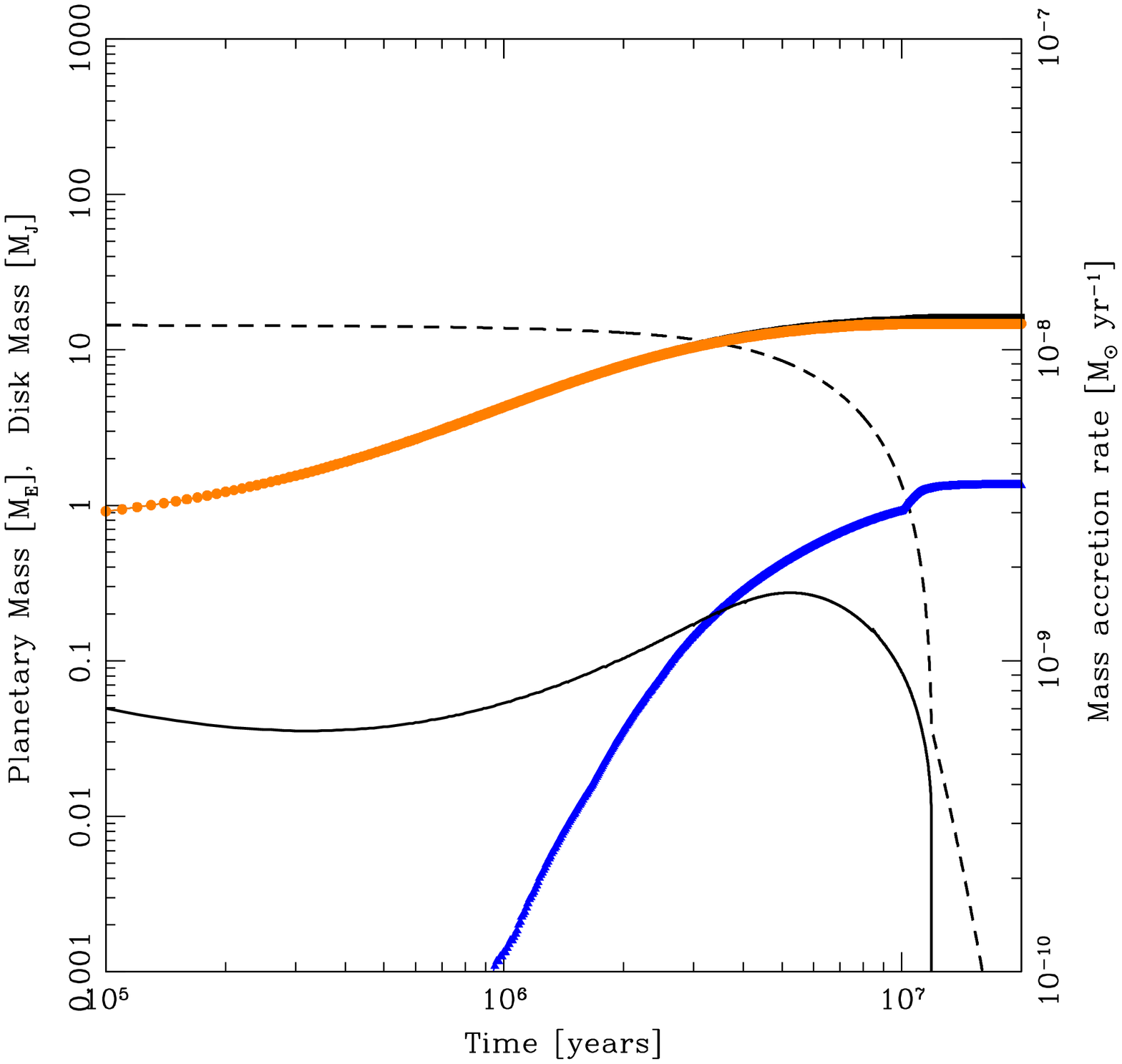}
%\includegraphics[width=7cm]{figs4/plmass_diskmass_5au_nomig_sigma1e3.eps}
%\unitlength1cm
%\begin{minipage}[t]{9cm}
%%\hspace{0.5cm}
%\begin{minipage}[h]{9cm}
%\begin{picture}(1,6)
%\includegraphics[width=4.5cm]{figs4/plmass_diskmass_diskaccr_5au_opa1_sigma1e3.eps}
%\end{picture}
%\hspace{3.5cm}
%\begin{picture}(7,6)
%\includegraphics[width=4.5cm]{figs4/plmass_diskmass_diskaccr_5au_opa003_sigma1e3.eps}
%\end{picture}
%\end{minipage}
%\end{minipage}
%\begin{minipage}[t]{9cm}
%
\caption[Mass evolution1]{Evolution of disk mass and accretion rate, in
presence of a DZ, and
%Mass evolution in a disk with a dead zone and
\(\Sigma_0=10^3 \ {\rm g \ cm^{-2}}\),
with standard opacity (top panel, RunE1) and reduced
opacity (bottom panel, RunE2). Also plotted are disk mass evolution (dashed
curves) along with the disk accretion rate onto the central star (thin
solid curves.) \label{fig9}}
%\end{minipage}
\end{figure}
\begin{figure}
%\epsscale{2}\plottwo{figs4/plmass_diskmass_diskaccr_5au_opa1_sigma1e4.eps}{figs4/plmass_diskmass_diskaccr_5au_opa003_sigma1e4.eps}
\epsscale{2}\plottwo{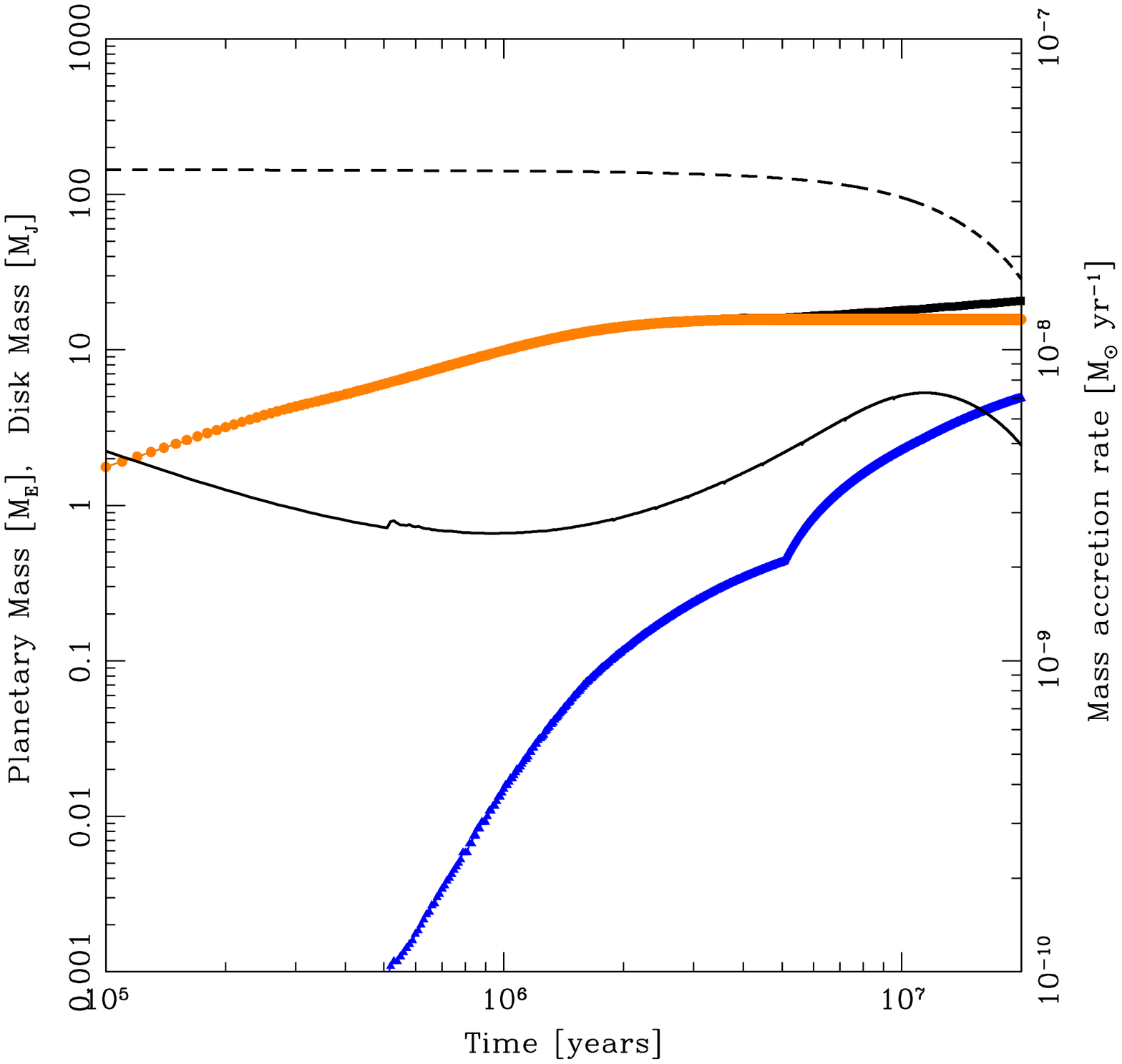}{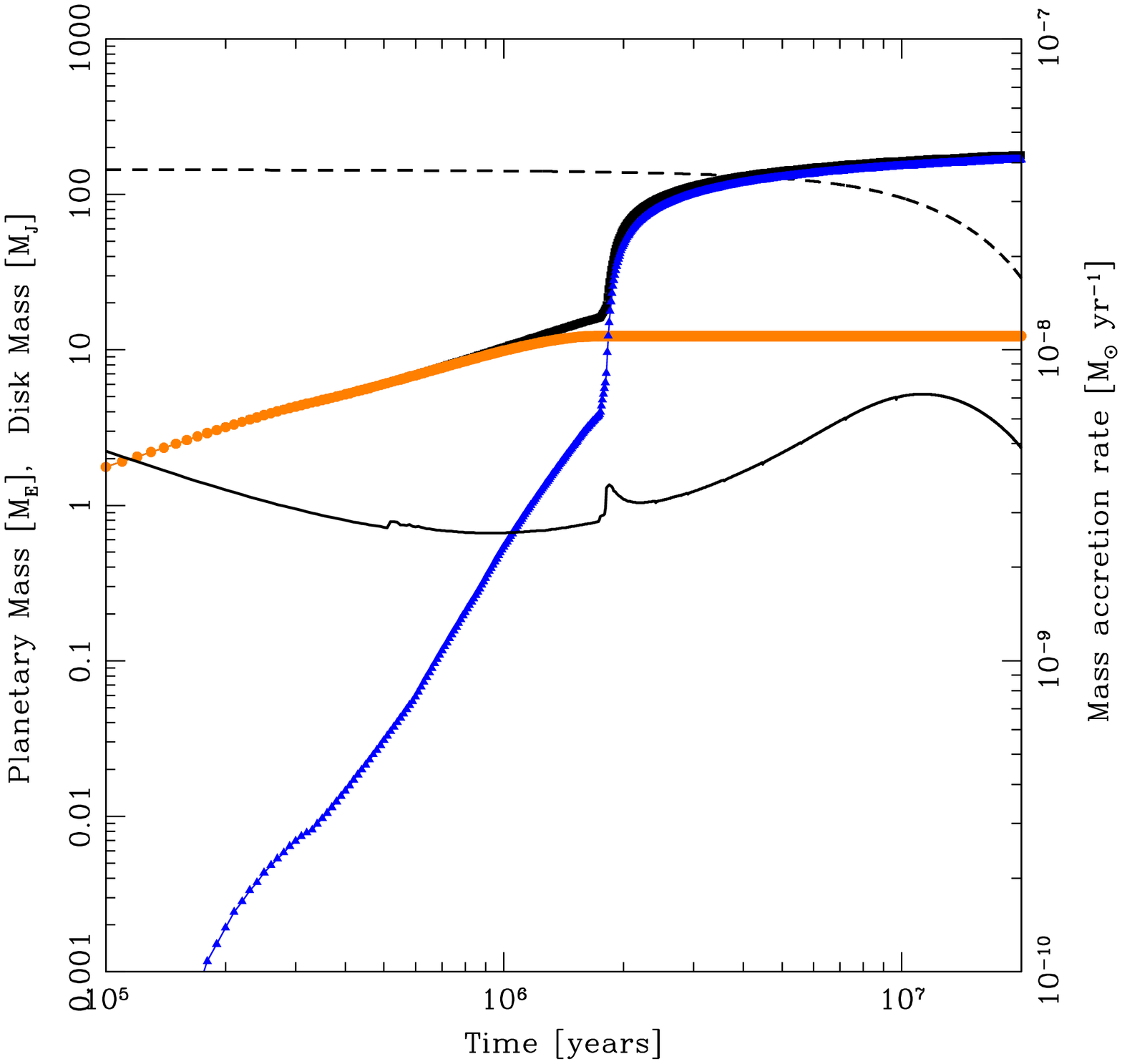}
%\includegraphics[width=7cm]{figs4/plmass_diskmass_5au_nomig_sigma1e4.eps}
%\unitlength1cm
%\begin{minipage}[t]{9cm}
%%\hspace{0.5cm}
%\begin{minipage}[h]{9cm}
%\begin{picture}(1,6)
%\includegraphics[width=4.5cm]{figs4/plmass_diskmass_diskaccr_5au_opa1_sigma1e4.eps}
%\end{picture}
%\hspace{3.5cm}
%\begin{picture}(7,6)
%\includegraphics[width=4.5cm]{figs4/plmass_diskmass_diskaccr_5au_opa003_sigma1e4.eps}
%\end{picture}
%\end{minipage}
%\end{minipage}
%\begin{minipage}[t]{9cm}
%
\caption[Mass evolution1]{Evolution of disk mass and accretion rate (as in
Fig. \ref{fig9}, except for column density
\(\Sigma_0=10^4 \ {\rm g \ cm^{-2}}\), with standard (top panel, RunF1)
and reduced (bottom panel, RunF2) opacity.
%Mass evolution in a disk with a dead zone and \(\Sigma_0=10^4 \ {\rm g \
%cm^{-2}}\) with a standard opacity (top panel) and a reduced
%opacity (bottom panel.) Also plotted are disk mass evolution (dashed
%curves) along with the disk accretion rate onto the central star (thin
%solid curves.)
\label{fig10}}
%\end{minipage}
\end{figure}
First, we study the case with no migration, but with disk evolution.
This, of course, is an unlikely scenario, because a growing
planetary core would quickly migrate toward the central star,
especially inside the dead zone due to an enhanced disk mass (see
MPT07).  We show these cases nevertheless, since they provide
reasonable estimates of planetary masses achievable in our fiducial
disks.

Fig. \ref{fig9} and \ref{fig10} show the results of
\(\Sigma_0=10^3\) and \(10^4 \ {\rm g \ cm^{-2}}\) respectively with
standard (top panels) and reduced (bottom panels) opacities (see \S
2.3 and Table \ref{tb1} for initial conditions for RunE1(2), and
RunF1(2)). The core has an initial mass of \(0.6 M_E\) as before,
and stays at 5 AU throughout the simulations. The planetesimal size
is 100 m. In all of these cases, the protoplanetary core is
originally inside the dead zone, and is left outside it as the disk
grows and the dead zone shrinks. The final planetary mass of the
case with a massive disk and a reduced opacity (bottom panel of Fig.
\ref{fig10}) is about half a Jupiter mass, while all the other cases
produce planets about a Neptune mass (\(20 M_E\)).
Note that in both MMSN-mass and more massive disks, the opacity reduction
leads to more efficient gas accretion as seen in \S 3.2.
These results further emphasize the importance of the opacity reduction in
helping rapid planet formation, and indicate
that it is difficult to form a Jupiter mass planet in a MMSN-mass disk.

Also plotted in Fig. \ref{fig9} and \ref{fig10} are evolution of
disk mass (dashed lines) and disk accretion rate onto the central
star (thin solid lines.)  As shown in \S 2.2, a dead zone enforces a
nearly constant accretion rate onto the star, and a rapid dispersal
of a disk once it's gone.
%
%This, of course, is an unlikely scenario, because a growing
%planetary core would quickly migrate toward the central star,
%especially inside the dead zone due to an enhanced disk mass (see
%MPT07). In such a case, unless the planet were intercepted somehow,
%for example by the inner edge of the dead zone
%\citep{Masset06,Morbidelli08}, the stellar magnetosphere
%\citep{Lin96}, or by the magnetic resonances \citep{Terquem03}, the
%(proto-)planet would be lost into the star. If the planet survived
%and were left behind the dead zone, the planet would follow the
%evolution of the dead zone, and accretes gas and planetesimals until
%it reaches the gap-opening mass. Even in this case, since the
%accretable amount of gas mass increases outward, it is unlikely for
%a planet to achieve mass larger than the values obtained above,
%unless the planet migrates outward.
%
%The dead zone initially extends out to \(13\) AU, and the disk
%viscosity parameter inside it is \(\alpha=10^{-5}\), while
%\(\alpha=10^{-2}\) outside it. The gas surface mass density is
%\(\Sigma_{gas}=1000(r/AU)^{-3/2}\), and the dust surface mass
%density is \(\Sigma_{solid}=300(r/AU)^{-2}\).
%
%We place an protoplanetary core of \(0.6 M_E\) both inside and
%outside the dead zone, and compare the results.
%just inside the dead zone (\(10\) AU) as well as just outside it (\(15\) AU) to
%compare their results with each other.
%
\subsection{With planet migration}
\begin{figure}
%\epsscale{2}\plottwo{figs4/disk_8au_sigma1e3.eps}{figs4/plmass_diskmass_diskaccr_8au_sigma1e3.eps}
\epsscale{2}\plottwo{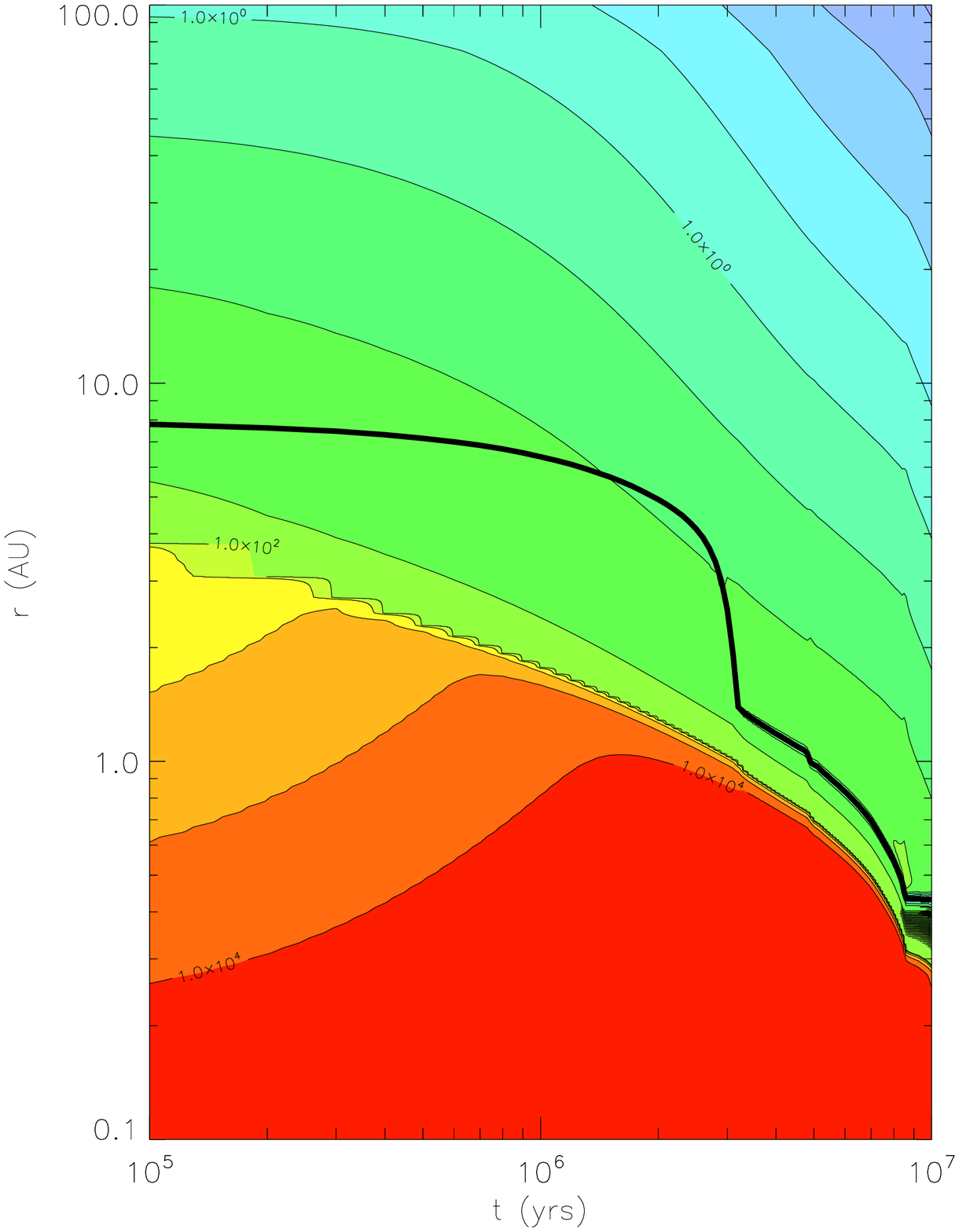}{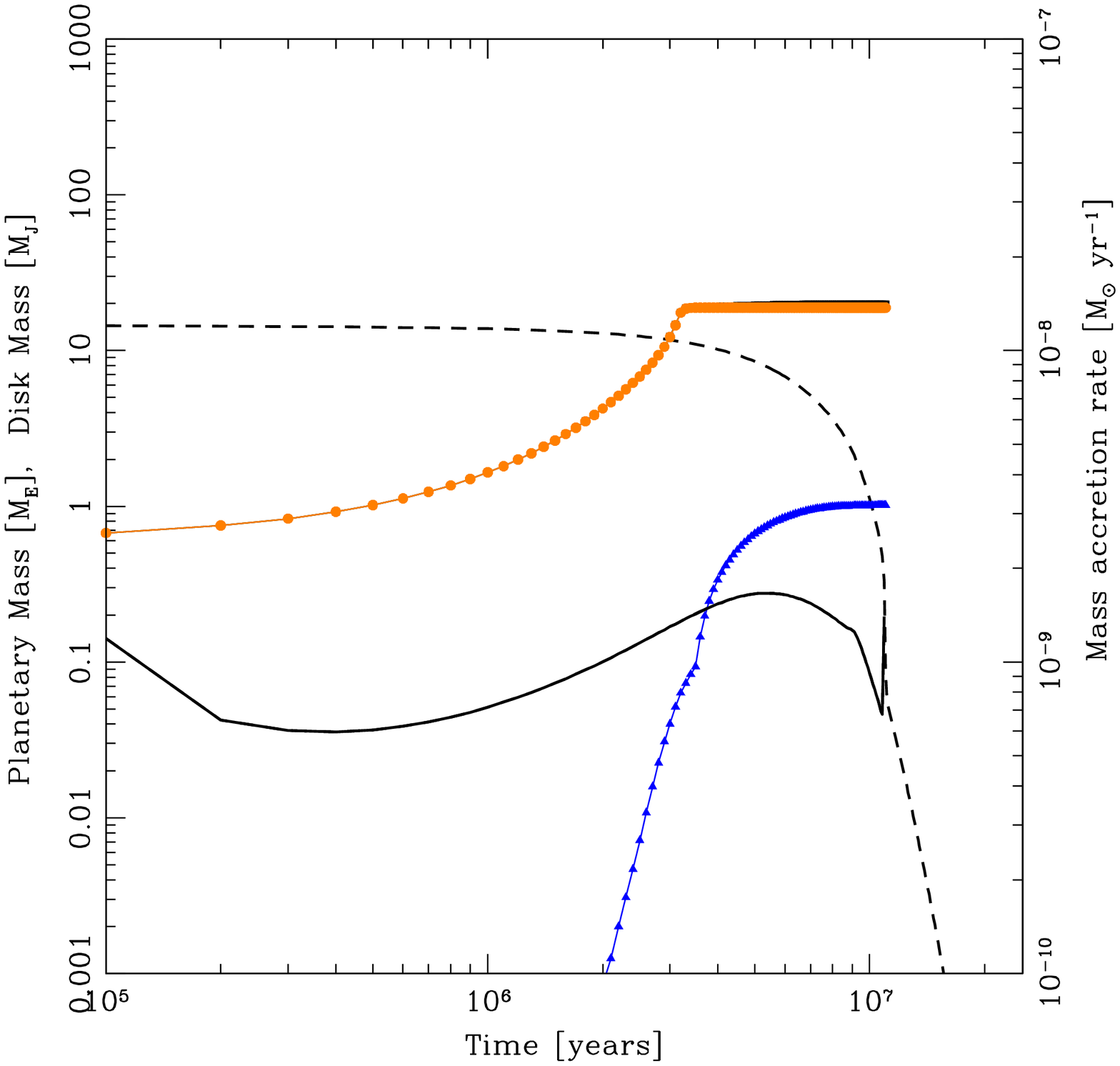}
%\unitlength1cm
%\begin{minipage}[t]{9cm}
%%\hspace{0.5cm}
%\begin{minipage}[h]{9cm}
%\begin{picture}(1,6)
%\includegraphics[width=4.5cm]{figs4/disk_8au_sigma1e3.eps}
%\end{picture}
%\hspace{3.5cm}
%\begin{picture}(7,6)
%\includegraphics[width=4.5cm]{figs4/plmass_diskmass_diskaccr_8au_sigma1e3.eps}
%\end{picture}
%\end{minipage}
%\end{minipage}
%\begin{minipage}[t]{9cm}
%%\vspace{0.5cm}
%
\caption[fig11]{Evolution of planet and disk with a planetary core initially
just inside DZ at 8 AU. The disk column density \(\Sigma_0=10^3 \ {\rm g \ cm^{-2}}\),
reduced opacity, and a 100 m-size planetesimal disk are assumed (RunG1).
Top: Evolution of both disk and planetary orbital radius. Note that the planet
appears as if it were initially outside the DZ, because the time axis is logarithmic and
starts at \(10^5\) yr.
%Disk evolution for \(\Sigma_0=10^3 \ {\rm g \ cm^{-2}}\) with a dead zone.
%A reduced opacity and a 100 m-size planetesimal disk are assumed. The planet
%is originally at 8 AU.
Bottom: Corresponding evolution of planetary
and disk masses as well as the disk accretion rate onto the central
star. \label{fig11}}
%\end{minipage}
\end{figure}
%
%\begin{figure}
%\unitlength1cm
%\begin{minipage}[t]{9cm}
%%\hspace{0.5cm}
%\begin{minipage}[h]{9cm}
%\begin{picture}(1,6)
%\includegraphics[width=4.5cm]{figs4/disk_9au_sigma1e3.eps}
%\end{picture}
%\hspace{3.5cm}
%\begin{picture}(7,6)
%\includegraphics[width=4.5cm]{figs4/plmass_diskmass_diskaccr_9au_sigma1e3.eps}
%\end{picture}
%\end{minipage}
%\end{minipage}
%\begin{minipage}[t]{9cm}
%%\vspace{0.5cm}
%
%\caption[fig12]{Top: Same as Fig. \ref{fig11},
%but the planet is originally at 9 AU.
%Bottom: Corresponding evolution of planetary and
%disk masses as well as disk accretion rate onto the central star. \label{fig12}}
%\end{minipage}
%\end{figure}
%
\begin{figure}
%\epsscale{2}\plottwo{figs4/disk_10au_sigma1e3.eps}{figs4/plmass_diskmass_diskaccr_10au_sigma1e3.eps}
\epsscale{2}\plottwo{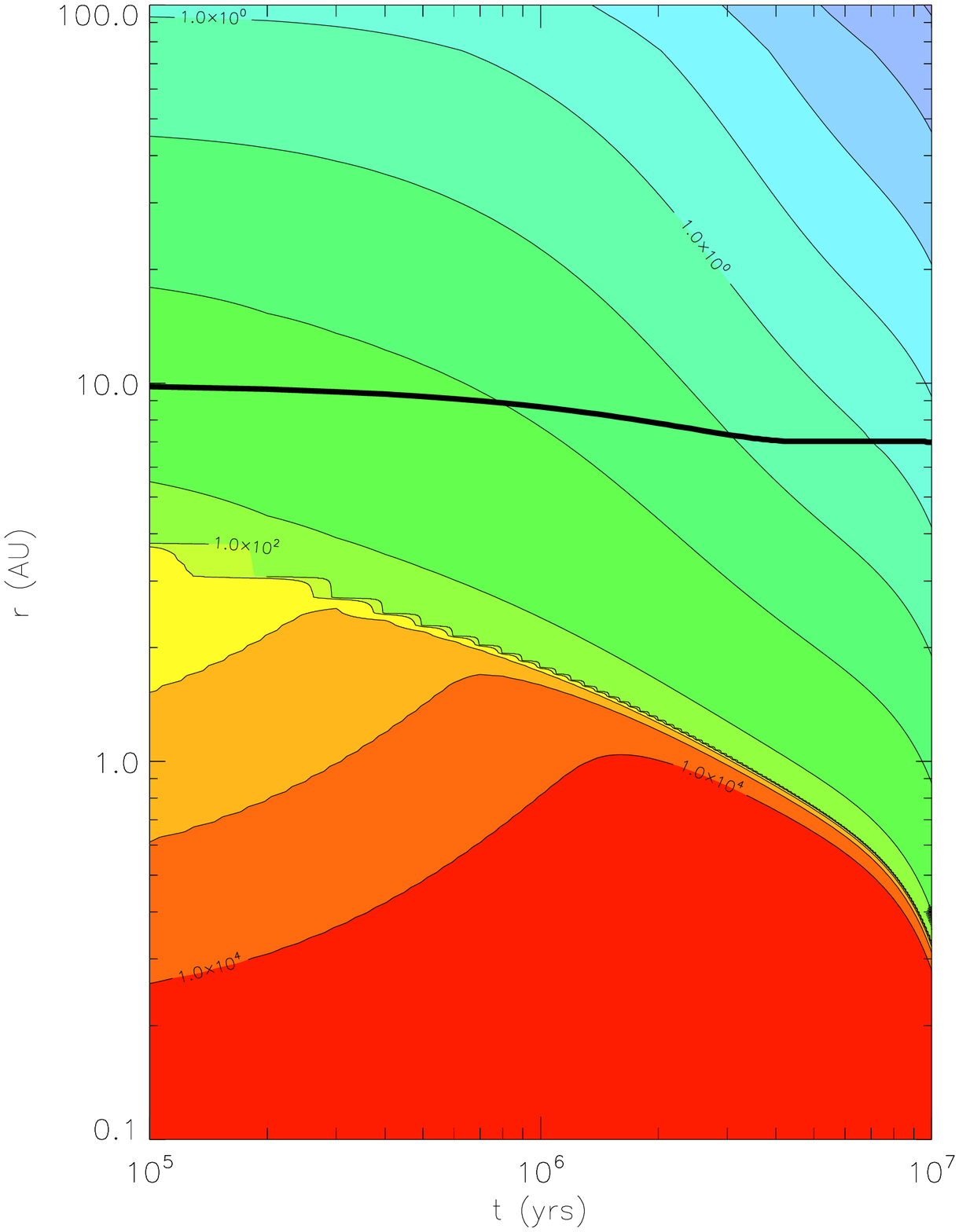}{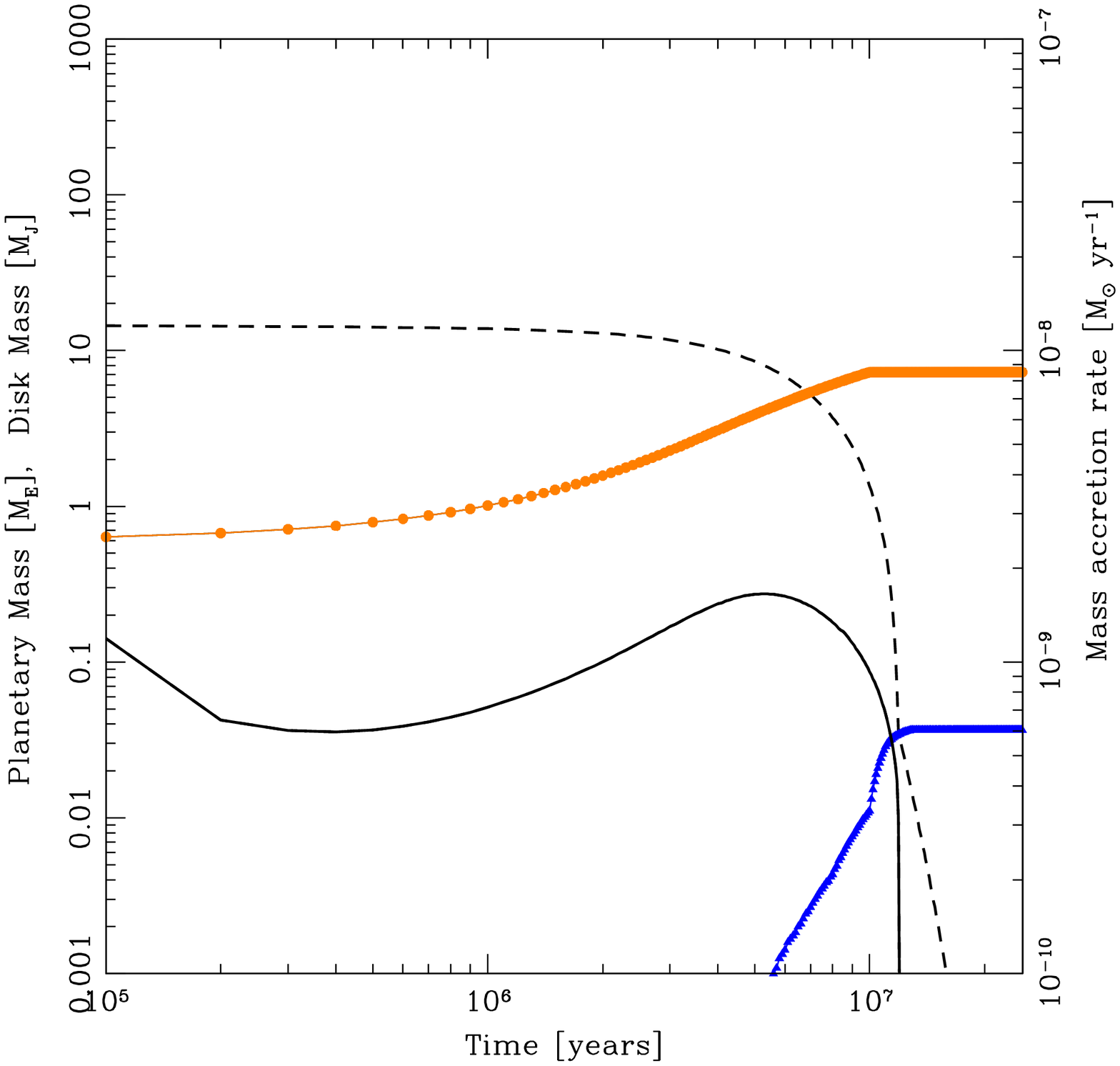}
%\unitlength1cm
%\begin{minipage}[t]{9cm}
%%\hspace{0.5cm}
%\begin{minipage}[h]{9cm}
%\begin{picture}(1,6)
%\includegraphics[width=4.5cm]{figs4/disk_10au_sigma1e3.eps}
%\end{picture}
%\hspace{3.5cm}
%\begin{picture}(7,6)
%\includegraphics[width=4.5cm]{figs4/plmass_diskmass_diskaccr_10au_sigma1e3.eps}
%\end{picture}
%\end{minipage}
%\end{minipage}
%\begin{minipage}[t]{9cm}
%%\vspace{0.5cm}
%
\caption[fig13]{Evolution of planet and disk as in Fig. \ref{fig11}, except the
initial orbital radius of a planet is outside DZ at 10 AU (RunG2).
%Top: Evolution of
%both disk and planetary orbital radius.  Bottom: Corresponding
%evolution of planetary and disk masses as well as the disk accretion
%rate onto the central star.
\label{fig13}}
%\end{minipage}
\end{figure}
Now, we finally show the cases with planetary formation and
migration in an evolving disk with a dead zone. The initial
conditions are the same as the previous subsection, except that
planetary migration is allowed with our fiducial softening parameter
$B=0.6$.

In Fig. \ref{fig11}, and \ref{fig13}, we show the evolution of a planetary
core initially at 8, and 10 AU in a MMSN-like disk with
\(\Sigma_0=10^3 \ {\rm g \ cm^{-2}}\), by assuming a reduced opacity and a 100 m-size
planetesimal disk (RunG1 and G2, respectively).
%We use a reduced opacity and a 100 m-size planetesimal disk.
Since the initial outer dead zone radius is $8.2$ AU,
the protoplanet starts inside(outside) the dead zone in RunG1(G2).
Note that, in Fig. \ref{fig11} (RunG1), the protoplanet appears as if it were
initially outside the dead zone, because the time axis is
logarithmic and starts at \(10^5\) yr.
In this case, the protoplanet is originally located inside the dead zone,
but is left outside it as the dead zone shrinks over time.
%a planet in Run G1(G2) starts inside(outside) the dead zone.
%both of these cores start {\it inside} the dead zone, but are left
%outside it as the dead zone shrinks.
%We use a reduced opacity and a 100 m-size planetesimal disk.

The core building phase takes longer as the initial orbital radius
moves outward, and as a result, the outer planet accretes less
amount of gas.  As expected from \S 5.1, neither case leads to a gas
giant. Final masses are \(20 M_E\) for 8 AU case (RunG1), and \(7
M_E\) for 10 AU case (RunG2). In RunG2, the core accretion is so
slow that the planet barely migrates compared to RunG1.  We obtain a
similar result to RunG1 for an initial radius just outside the dead
zone at 9 AU.

From the comparison of Fig. \ref{fig11} with Fig. \ref{fig7}, the
effect of a dead zone is apparent. Instead of losing a growing
protoplanet to the central star, the dead zone stops the migration
of the protoplanet by balancing the inner and outer torques. As a
result, the protoplanet has a chance of growing further at a slower,
viscous accretion rate of the disk.

%\begin{figure}
%\epsscale{2}\plottwo{figs4/disk_10au_270_sigma1e4.eps}{figs4/plmass_diskmass_diskaccr_10au_270_sigma1e4.eps}
%%\plottwo{fig14a.eps}{fig14b.eps}
%%\unitlength1cm
%%\begin{minipage}[t]{9cm}
%%%\hspace{0.5cm}
%%\begin{minipage}[h]{9cm}
%%\begin{picture}(1,6)
%%\includegraphics[width=4.5cm]{figs4/disk_10au_270_sigma1e4.eps}
%%\end{picture}
%%\hspace{3.5cm}
%%\begin{picture}(7,6)
%%\includegraphics[width=4.5cm]{figs4/plmass_diskmass_diskaccr_10au_270_sigma1e4.eps}
%%\end{picture}
%%\end{minipage}
%%\end{minipage}
%%\begin{minipage}[t]{9cm}
%%%\vspace{0.5cm}
%%
%\caption[fig14]{Evolution of plnaet and disk as in Fig. \ref{fig11}, except
%disk column density of \(\Sigma_0=10^4 \ {\rm g \ cm^{-2}}\) (RunH1).
%Note that the protoplanet appears as if it were initially outside the DZ,
%because the time axis is logarithmic and starts at \(10^5\) yr.
%%Top: Evolution of both disk and planetary orbital radius.
%%Disk evolution for
%%\(\Sigma_0=10^4 \ {\rm g \ cm^{-2}}\) with a dead zone. A reduced
%%opacity and a 100 m-size planetesimal disk are assumed. The planet
%%is originally at 10 AU.
%%Bottom: Corresponding evolution of planetary
%%and disk masses as well as the disk accretion rate onto the central
%%star.
%\label{fig14}}
%%\end{minipage}
%\end{figure}
%
\begin{figure}
%\epsscale{2}\plottwo{figs4/disk_15au_270_sigma1e4.eps}{figs4/plmass_diskmass_diskaccr_15au_270_sigma1e4.eps}
\epsscale{2}\plottwo{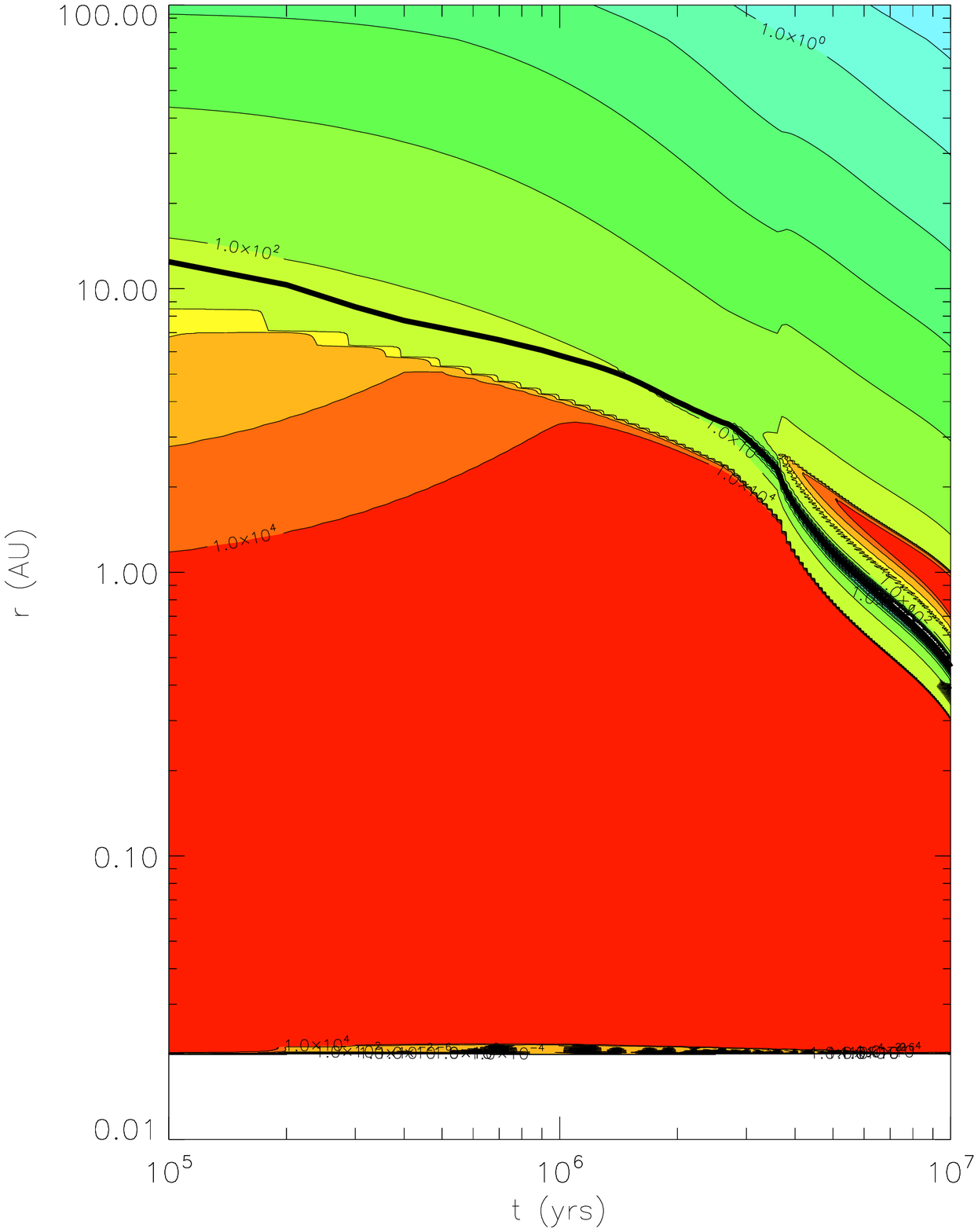}{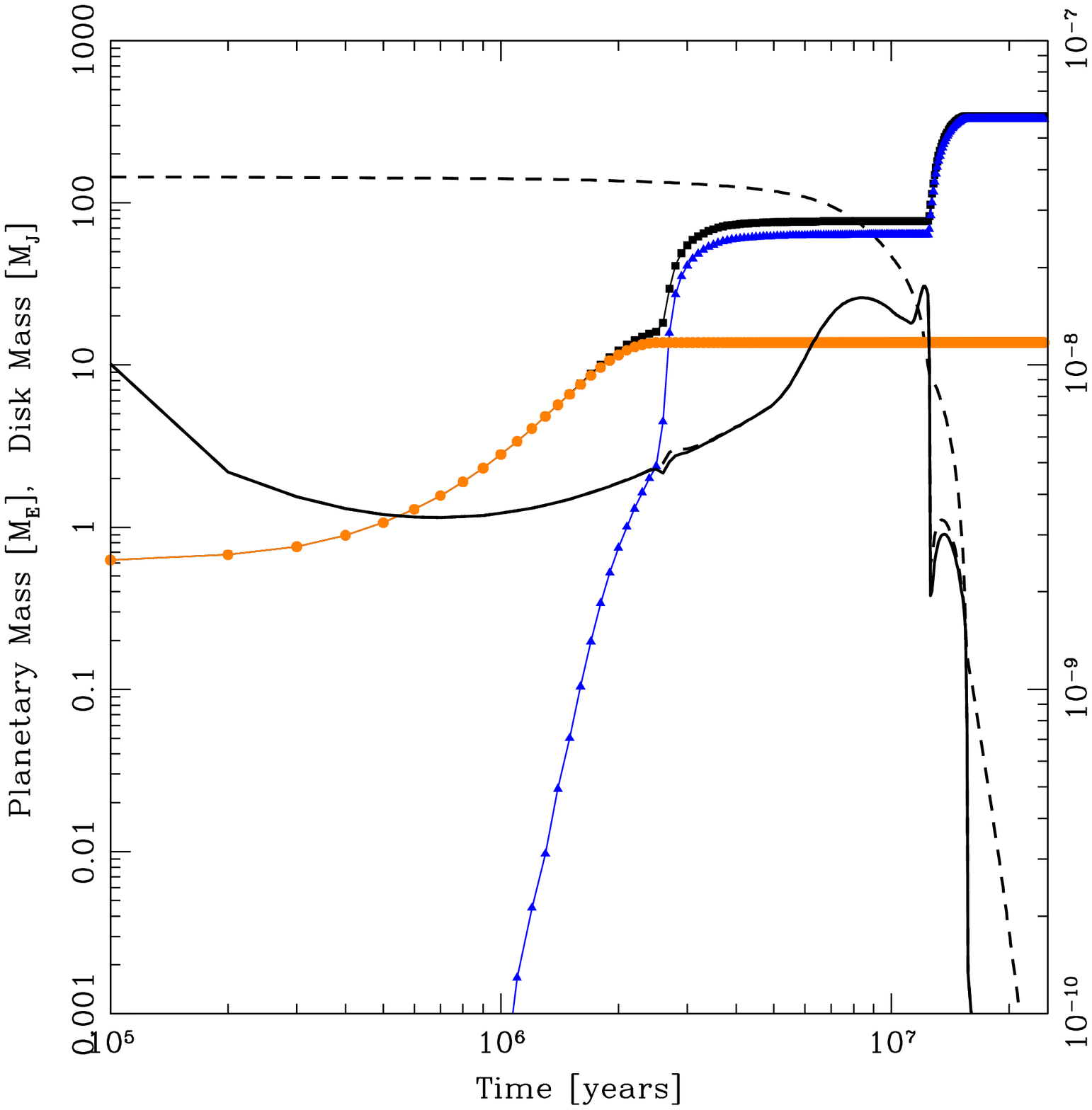}
%\unitlength1cm
%\begin{minipage}[t]{9cm}
%%\hspace{0.5cm}
%\begin{minipage}[h]{9cm}
%\begin{picture}(1,6)
%\includegraphics[width=4.5cm]{figs4/disk_15au_270_sigma1e4.eps}
%\end{picture}
%\hspace{3.5cm}
%\begin{picture}(7,6)
%\includegraphics[width=4.5cm]{figs4/plmass_diskmass_diskaccr_15au_270_sigma1e4.eps}
%\end{picture}
%\end{minipage}
%\end{minipage}
%\begin{minipage}[t]{9cm}
%%\vspace{0.5cm}
%
\caption[fig15]{Evolution of planet and disk as in Fig. \ref{fig11}, except
disk column density of \(\Sigma_0=10^4 \ {\rm g \ cm^{-2}}\), and
the initial orbital radius of a planet just inside DZ at 15 AU (RunH1).
Note that the protoplanet appears as if it were initially outside the DZ,
because the time axis is logarithmic and starts at \(10^5\) yr.
The second jump in mass seen on the bottom panel occurs at
semi-major axis smaller than our resolution limit,
and may not represent the true evolution of the planet.
%Top: Same as Fig. \ref{fig14},
%but the planet is originally at 15 AU. Bottom: Corresponding
%evolution of planetary and disk masses as well as the disk accretion
%rate onto the central star.
\label{fig15}}
%\end{minipage}
\end{figure}
\begin{figure}
%\epsscale{2}\plottwo{figs4/disk_15au_270_opa1_sigma1e4.eps}{figs4/plmass_diskmass_diskaccr_15au_270_opa1_sigma1e4.eps}
\epsscale{2}\plottwo{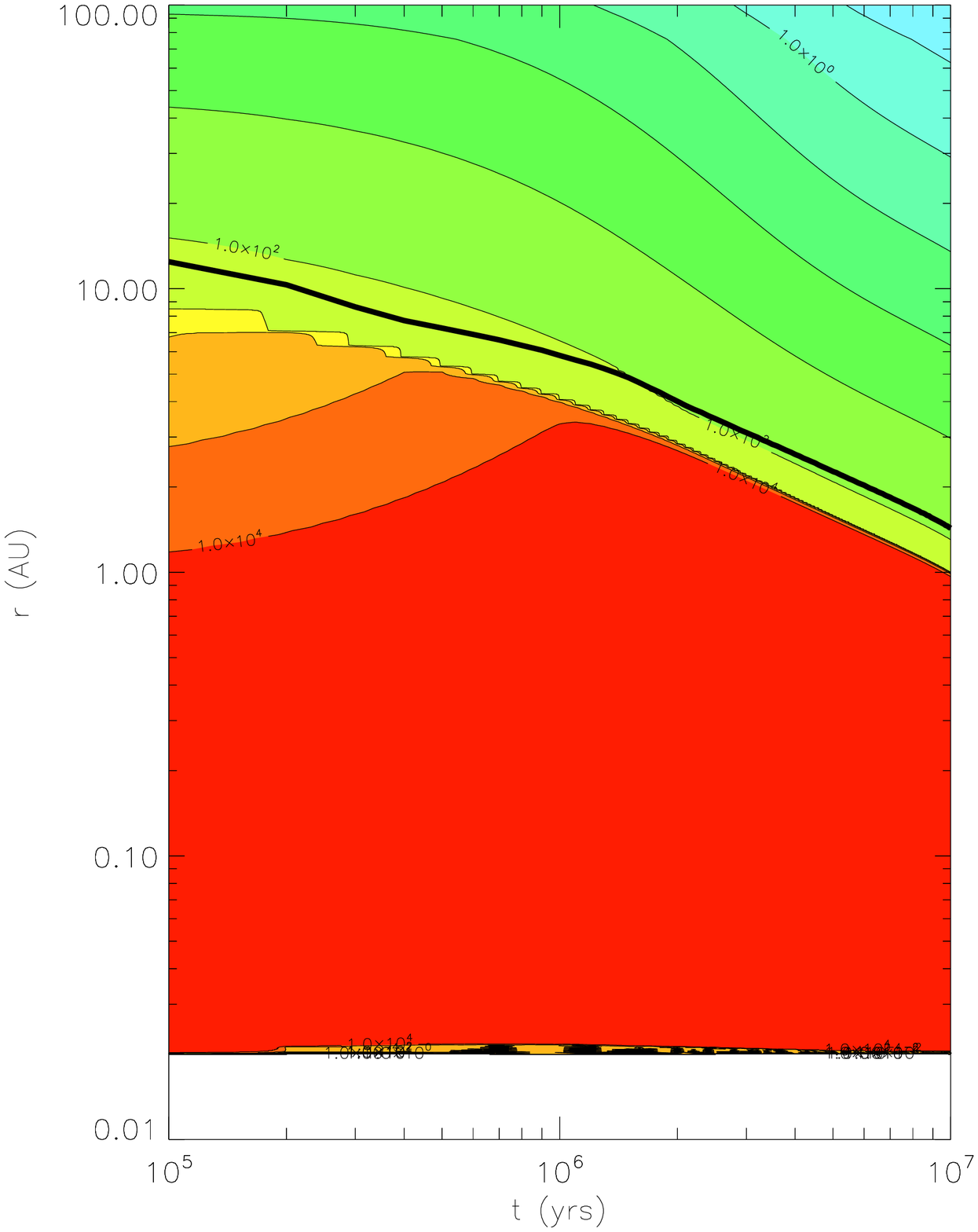}{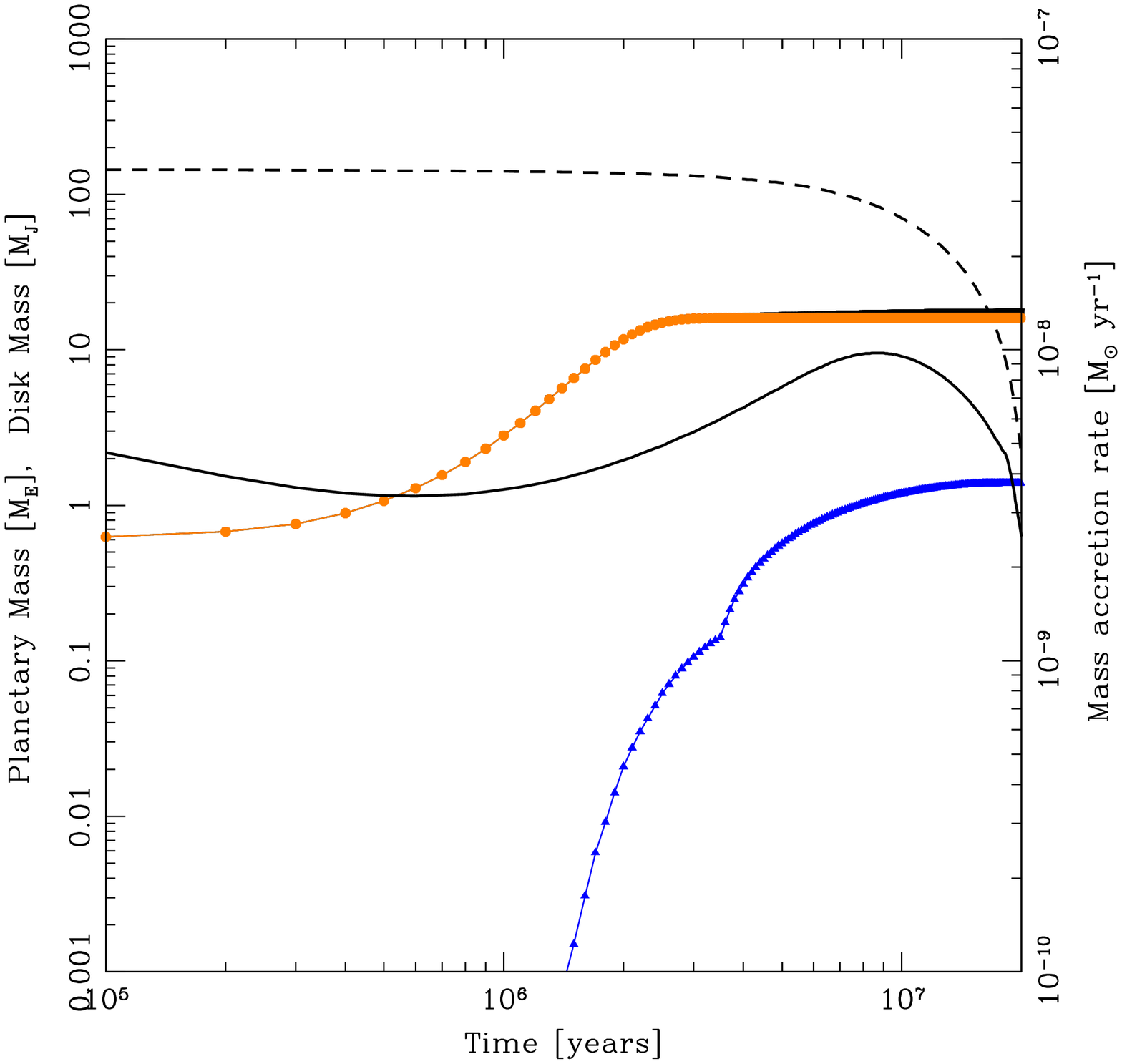}
%\unitlength1cm
%\begin{minipage}[t]{9cm}
%%\hspace{0.5cm}
%\begin{minipage}[h]{9cm}
%\begin{picture}(1,6)
%%\includegraphics[width=4.5cm]{figs4/disk_20au_270_sigma1e4.eps}
%\includegraphics[width=4.5cm]{figs4/disk_15au_270_opa1_sigma1e4.eps}
%\end{picture}
%\hspace{3.5cm}
%\begin{picture}(7,6)
%%\includegraphics[width=4.5cm]{figs4/plmass_diskmass_diskaccr_20au_270_sigma1e4.eps}
%\includegraphics[width=4.5cm]{figs4/plmass_diskmass_diskaccr_15au_270_opa1_sigma1e4.eps}
%\end{picture}
%\end{minipage}
%\end{minipage}
%\begin{minipage}[t]{9cm}
%%\vspace{0.5cm}
%
\caption[fig16]{Evolution of planet and disk as in Fig. \ref{fig15}, but
the standard opacity instead of reduced opacity is assumed (RunH2).
Again, although the initial protoplanetary radius is just inside DZ at 15 AU,
the protoplanet appears as if it were initially outside the DZ, because
the time axis is logarithmic and starts at \(10^5\) yr.
%Results for RunH3. Top: Same as Fig. \ref{fig15},
%but the standard opacity is assumed. Bottom: Corresponding evolution
%of planetary and disk masses as well as the disk accretion rate onto the
%central star.
\label{fig16}}
%\end{minipage}
\end{figure}
%
%Fig. \ref{fig14}, and \ref{fig15} compare the cases with a planetary
%core initially at 10, and 15 AU
Fig. \ref{fig15}, and \ref{fig16} show the evolution of a planetary
core initially at $15$ AU in a more massive disk with
\(\Sigma_0=10^4 \ {\rm g \ cm^{-2}}\), by assuming a reduced
(RunH1), and a standard (RunH2) opacity, respectively. For both
cases, a 100 m-size planetesimal disk is used. Since the initial
outer dead zone radius is $15.8$ AU, these planets start just {\it
inside} the dead zone. However, the protoplanets appear as if they
were initially outside the dead zone, because the time axis is
logarithmic and starts at \(10^5\) yr.
%Similar to the previous examples, the core building phase
%takes longer for the further initial radius as expected.
%%But the effect is not as visible since gas accretion is more efficient.
Run H1 obtains a gas giant with \(\sim 0.3 M_J\) in roughly 2.5 Myr,
while in Run H2, the protoplanet grows only up to $\sim 20 M_E$.
%migrates toward the star and is removed out of the system by 1 Myr.
%
%In the bottom panel of Fig. \ref{fig14}, mass accretion rate drops
%suddenly when a planet plunges into the star, because the disk opens
%a transient ``inner hole''.
%
%It should be noted that the resolution of our simulation is good
%down to ~0.1 AU, and that we don't define the inner dead zone edge.
%Therefore, any planets which appear to be ``lost'' in our
%simulations may still be alive if we improve our resolution, take
%account of the inner dead zone edge, or include the star-planet
%tidal interaction.
%
%Fig. \ref{fig16} is a similar run to Fig. \ref{fig15}, but with a
%standard opacity. In this case, the planet grows only up to $\sim 20 M_E$.
This demonstrates that we need an opacity reduction (or some other
mechanism) even when we include planet migration. As emphasized in
\S 4, this is the result of a slower oligarchic core accretion
compared to a runaway core accretion. The overall mass accretion
onto a protoplanet does not speed up as a result of migration, and 
hence the constant replenishment of the planetesimal feeding
zone.

Also, by comparing the bottom panels of Fig. \ref{fig15} and
\ref{fig16}, it appears that a gap-opening planet speeds up the
dispersal of a gas disk.  This additional disk dispersal mechanism
should be investigated in multiple planet formation models.
%
%
%--- Section 5 ------------------------------------------------------------------------------
%
\section{Discussion and Conclusions}
We have presented some new results on planet formation and migration
in evolving disks with dead zones.  The most significant is that
dead zones provide a natural way of saving planetary systems even as
the planets migrate through disks whose properties change
significantly over hundreds of thousands to millions of years. The
dissipation of the disk does place interesting constraints on their
masses - we find that only Neptunian mass planets can be formed in
the MMSN-mass disk models.  Jovian planet formation requires more
massive disks, which has also been suggested by other groups (e.g.
P96 and HBL05), and was recently demonstrated by multiple planet
formation simulations by \cite{Thommes08}. The time scale that we
find for Jovian planet formation in these more massive disks - about
2.5 Myr - is within observational limits of disk lifetimes. While
not drastically reduced from previous estimates, our Jovian planet
formation time incorporates many new aspects including migration in
the presence of dead zones, effects of slower oligarchic growth that
are more realistic than earlier accretion scenarios, and an effect
of a reduced opacity.

Dead zones turn out to play a potentially important role as
blockades against inward planetary motion.  Our planetary cores are
generally kept outside of the dead zone - and therefore immersed in
the region where they can ``feed'' more effectively on surrounding
gas.  When the planets become massive enough to open a gap, the
outwards directed torque associated with the density gradient of the
outer edge of the dead zone weakens, and the planets migrate into
the dead zone. Planets migrate slowly inside the dead zone due to
the lower viscosity there compared to outside it.

The methods we used were straightforward to implement in our
planetary/disk evolution code. In order to model dead zone
evolution, we simply assumed that the dead zone is where the surface
mass density is above the critical value
(\(\Sigma>\Sigma_{crit}=21\), and \(80 \ {\rm g \ cm^{-2}}\) for
\(\Sigma_0=10^3\) and \(10^4 \ {\rm g \ cm^{-2}}\) respectively)
that was determined from a stationary disk model in MP06.  Also, we
have included gas accretion through the surface layers, which
contributes to make a dead zone shrink very rapidly (see Fig.
\ref{fig3}). Combining these two conditions, we have found that the
dead zone radius shrinks from \(8.2\) AU to \(2\) AU within 2 Myr 
for \(\Sigma_0=10^3 \ {\rm g \ cm^{-2}}\).

For the planet formation part of the code, we follow the approach by
P96, but with a few differences.  First of all, we assume that the
planetesimal accretion stage is mainly carried out by oligarchic
growth, rather than by rapid runaway accretion used by P96. This is
a reasonable assumption since the runaway growth most likely ceases
long before the planetary mass reaches a few times \(10^{-5} M_E\)
\citep{Thommes03}, which is much smaller than our initial planetary
mass (\(0.6 M_E\)), and switches to a slower oligarchic growth
\citep{Ida93,Kokubo98}. Depending on the size distribution of
planetesimals, the rate of accretion could be significantly different.
%Since we built our formation model based on the
%results obtained by P96, this difference between planetesimal
%accretion rates s that we choose a rather high surface mass density
%for planetesimals (\(\Sigma_{solid}=1000(r/AU)^{-2}\)).
Secondly, we parameterize our gas accretion rate following INE00 and
\cite{DAngelo03}.
%so that we can reproduce the results by P96 (see \S 2.2 as well).
Therefore, we are not calculating planetesimal and gas accretion
rates in an interactive way as in P96, HBL05, or A05. However, our
simulations show reasonable agreements with their results (see \S
3). Also, since we take account of the subdisk accretion phase, the
final stage of gas accretion slows down, rather than exponentially
increases as in P96, or artificially cuts off as in HBL05. Thirdly,
we include a 1D gas disk evolution and planet migration.  Our
approach is similar to A05, but we don't include photoevaporation
effects for this study, since this is likely to be negligible during
planet formation.
%dead zone helps a natural, rapid disk dispersal.
%In other words, we cut off the planetesimal accretion artificially, and turn
%on the gas accretion when the core mass reaches \(10 M_E\).
%In P96, the core accretion slows down automatically due to the
%depletion of planetesimals in the feeding zone.
%Thirdly, the transition from the slow gas accretion to rapid gas
%accretion is set by the condition: \(r_{\rm Bondi}\sim h \sim r_{\rm
%Hill}\), while this transition naturally occurs in P96 as the core
%mass becomes comparable to the gas envelope's mass.  This assumption
%is reasonable, because beyond this point, all gas within the sphere
%with a radius equal to the pressure scale height is gravitationally
%dominated by the planet rather than the central star (\(r_{\rm
%Hill}>h\)), and the gas is bound to the planet rather than moving
%freely (\(r_{\rm Bondi}>h\), i.e. the gravitational energy is larger
%than the thermal energy). Therefore, the mass accretion occurs
%nonlinearly, and very quickly. The typical crossover mass that can
%be estimated from the Hill condition is \(\sim 40 M_E\).
%
%{\bf We should think how the density jump changes when it becomes
%Rayleigh unstable. Also corotation??}
%

There are several aspects of planet formation that we have not included
explicitly in this study. We list several below, but note that we do
not expect their absence to strongly affect our results.

First of all, our simple torque prescription does not capture the
nature of a turbulent disk properly. A number of numerical
simulations have shown that turbulent fluctuations can cause torque
fluctuations \citep[e.g.][]{Laughlin04,Nelson04}, which leads to
stochastic type I migration \citep{Johnson06}. This results in even
{\it faster} migration for most planets, but also allows {\it
outward} migration for some of them \citep{Johnson06}. The inclusion
of such an effect may allow a planet outside the dead zone to ``jump
the barrier'' provided by the edge of a dead zone. Even within a
dead zone, turbulent fluctuations in upper and lower active layers
may be able to affect planet migration significantly.

Also, we do not take account of the evolution of temperature profile
of a gas disk. To better evaluate the disk evolution, we should
include the radiative transfer to the code. This would affect not
only the dead zone and disk evolution, but also planet migration. A
recent work of \cite{Paardekooper08} demonstrated that, in a
non-isothermal disk, planet migration is preferentially outward at
around 5 AU until the disk mass decreases significantly. Therefore,
more precise treatment of a disk temperature may save type-I
migrators effectively, even inside a dead zone.

Perhaps the most important simplification of our model is the 1D
treatment of the disk, and the rather crude prescriptions for the
dead zone, which led to a very sharp density transition at the outer 
dead zone radius (see Fig. \ref{fig2}). This feature enabled 
a planet to stop its migration by
balancing inner and outer torques.  However, such a sharp density
gradient makes a disk locally Rayleigh unstable (i.e. an epicyclic
frequency $\kappa^2<0$), which may render the sharp transition from
the dead zone to the outer active disk much smoother. Even when the
disk is still locally stable, the Rossby wave instability can smooth
the density gradient in a similar manner when pressure varies
significantly over a few times the disk thickness \citep[][and the
references therein]{Li01}. This possibility should be checked
carefully by numerical simulations.

We did not take account of the effect of the corotation torque
either. This is a safe assumption most of the time, since the
corotation torque depends on the gradient of the inverse of the
specific vorticity ($\partial (\Sigma /B)/
\partial r$) \citep{Goldreich80,Masset01},
and therefore its effect becomes particularly important when the
surface mass density changes sharply, e.g. at the edge of a dead
zone. However, we do not expect this affects our results
significantly, because inner and outer Lindblad torques balance with
each other far away from the density jump, where the corotation
torque is likely negligible, and planet migration stalls (also see
Appendix B in MPT07). When Rayleigh, or Rossby wave instability
makes the density gradient at the outer edge of a dead zone
smoother, the corotation torque can become important. In such a
case, a planet would not be stopped at the outer edge of a dead
zone, but pulled into it instead.

The corotation torque is also found to be important for migration of
sub-Jovian mass planets like Saturn \citep{Masset03}.
% affect planet migration for sub-Jovian mass planets like Saturn \citep{Masset03}.
In a standard disk, the time scale of the so-called type III
migration for such planets is comparable to, or even shorter than, 
type I migration \citep{Masset03}. Although this is most relevant to
our RunH1 (Fig. \ref{fig15}), we don't expect a significant change
in our result. This is because type III migration is likely to
switch to slower type II migration as soon as the planet enters a
dead zone, where planets tend to open a wider gap at smaller mass
due to its low viscosity.

Another simplifications is that we have only considered a
protoplanetary disk with a single core, while the real disks are
expected to have multiple cores.  Studies of this kind have been
done by \cite{Chambers06,Thommes08} for disks with no dead zone, and
by \cite{Morbidelli08} at the {\it inner} edges of disks with dead
zones. A recent paper by \cite{Ida08ap}, took a similar approach to
us, and studied a dead zone's effects on retaining icy grains, as
well as protoplanetary cores. They reproduced the observed frequency
and mass-period distribution of gas giants around solar-type stars
with a moderate reduction in type I migration speed.
%{\bf Write a bit more...}
%It would be interesting to study such an
%interaction, and its effect on planet formation and migration.

Yet another simplification is that all of our planets in this study
are on circular orbits. Planet-disk interaction has been proposed as
a method to drive planetary eccentricity \citep{Goldreich03}, and
recent numerical simulations confirmed this for massive ($>M_J$)
planets \citep[e.g.][]{Masset04,DAngelo06}.
%the overall effect may be dominated by eccentricity damping
%Our results would be modified if we include the eccentricity effect.
%This is also interesting in relation to the former point, because
%the eccentricity could be enhanced either via planet-planet
%interaction \citep[e.g.][]{Rasio96}, or via planet-disk interaction
%\citep[e.g.][]{Goldreich03}.
If a planetary eccentricity is enhanced significantly (\(e\gg
h/r\)), then planet migration could be slowed down
\citep[e.g.][]{Papaloizou00,Papaloizou02}, and mass growth rate
could increase \citep[e.g.][]{DAngelo06,Kley06}.
%
%Another point which has not been included explicitly in our study is
%the variety of planetesimal sizes.  We have assumed a single size of
%\(10\) km for planetesimals, and as a result, used a rather heavy
%disk model (\(\Sigma_{solid}=1000(r/AU)^{-2}\)). However,
%\cite{Rafikov04} suggested that the protoplanetary cores could be
%assembled quickly by considering smaller planetesimals (\(\leq
%0.1-1\) km), because their random velocities excited by the cores
%are damped by gas drag.  If about \(\sim 10 \%\) of the solid mass
%is in these small planetesimals, his estimate shows that the
%protoplanetary core of \(10 M_E\) can be assembled within roughly
%\(10^5-10^6\) years at \(1-10\) AU for a minimum mass solar nebula
%model.
%
%We will leave these investigations to future work.

We also employ simplified planetesimal disks.  First, we only
consider single-size planetesimals ranging from 100 m to 100 km,
while the real disks should have multiple-size planetesimals.
Secondly, we don't include the planetesimals' effect on planet
migration, which could potentially be important \citep{Murray98}.
Thirdly, we assume the dispersion-dominated random velocities for
planetesimals with all sizes.  Since small planetesimals experience
strong gas drag and achieve reduced random velocities, their mass
accretion rate should be treated in the shear-dominated regime
\citep{Rafikov04,Chambers06}.  This may be important for 100 m-size
planetesimals \citep{Chambers06}, and the planetesimal accretion
would be runaway, rather than oligarchic in such a case.

Finally, we don't include the effects of
photoevaporation \citep[e.g.][]{Shu93,Hollenbach94}, which is likely
to be important during the last stage of disk evolution
\citep{Clarke01,Alexander06b}. However, photoevaporation becomes
important only when disk mass becomes significantly low (\(\sim 1
M_J\) or so), and therefore it is unlikely to affect our results.

We conclude by listing our major findings: \\

(1) Dead zones strongly evolve over the duration of the disk,
starting with outer radii of order $10-15$ AU, and shrinking with time to of
order an AU or less.  \\

(2) Protoplanets which are left outside the dead zone tend to
migrate inward as they accrete planetesimals, and stop just outside the
outer dead zone radius due to a steep surface mass density jump.
%and grow there.
Then such protoplanets migrate at the viscous time scale of the shrinking dead zone
until they achieve gap-opening mass.  Once they open a gap, they enter the
dead zone, and migrate inwards rather slowly due to the low viscosity there. \\
%the gap-opening mass is reached. \\

(3) The final mass of a planet is determined by disk mass.  In a
minimum-mass solar nebula disk, it is difficult to get a planet more
massive than Neptune.  We find Jovian planets form in disks that are
ten times more massive than this. \\
%, which has implications for submillimeter surveys for disk masses. \\

(4) A Jupiter mass planet can form within \(\sim 2.5\) Myr (see Fig.
\ref{fig15}) in a disk with a dead zone, by assuming standard type I
migration. However, since planet migration does not help core
accretion as much in oligarchic growth as in runaway growth
scenario, we expect that the opacity reduction (or some other mechanism) is
necessary to form a Jovian planet within a disk life
time. \\

(5) Dead zones may help explain the existence of long-lived
strong accretors. When a dead zone is present, the mass accretion
rate is likely to take a nearly constant value until the dead zone
disappears, rather than decreasing gradually as expected in a standard disk without
a dead zone. Once this happens, the rest of the disk is dispersed on
a few Myr time scale, unless photoevaporation and/or planet
formation provide additional sources of disk dissipation.
\\
%
%
%(2) Protoplanets which stay inside the dead zone tend to be left
%outside the dead zone as it shrinks. These planets follow a similar
%fate to the above case
%(see Fig. \ref{test924}). \\
%
%(3) Protoplanets which start migrating far from the star do not
%accrete enough planetesimals, and therefore do not form massive
%planets (see Fig. \ref{fig13}). \\
%
%(4) We did not see the effect of decreasing opacity on planet
%formation timescale. \\
%
%(5) By mimicking the 3D effect of the torque, the migration slows
%down as expected from \cite[e.g.][]{Tanaka02,Menou04}.  Including
%this effect, we see that...
%
\section*{Acknowledgments}
We thank the anonymous referee for useful comments. Our simulations
were carried out on computer clusters of the SHARCNET HPC Consortium
at McMaster University. S.~M. is supported by McMaster University, 
and Northwestern University. R.~E.~P. is supported by a grant from
the National Science and Engineering Research Council of Canada
(NSERC). E.~W.~T. is supported by a grant from NSERC, and during
earlier parts of this work, by the Spitzer Space Telescope
Theoretical Research Program and Northwestern University.
%
%--------------------------------------------------------------------------------------------
\bibliography{REF}
\bibliographystyle{apj}
%\bibliographystyle{mn2e}
%
%
%\label{lastpage}
%
\end{document}